\pdfoutput=1
\documentclass[letterpaper, 10pt]{article}
\usepackage[letterpaper,margin=1in]{geometry}

\usepackage{epsfig}
\usepackage{graphicx}
\usepackage{subfigure}
\usepackage{amsmath,amssymb}
\usepackage{color}
\usepackage{listings}
\usepackage[normalem]{ulem}
\usepackage{floatrow}
\usepackage{xspace}

\newcommand{\ie}{{\em i.e.,~}}
\newcommand{\eg}{{\em e.g.,~}}
\def\F{Figure~}
\def\Alg{Algorithm~}

\newcommand{\ar}[3]{} %% something bibtex is doing - ignore it
\include{latexdefs}
\include{mathnotation}

\definecolor{grey}{rgb}{0.5,0.5,0.5}

\renewcommand{\S}{Section~}

\newenvironment{proof-sketch}{\noindent{\bf Sketch of Proof}\hspace*{1em}}{\qed\bigskip}
\newenvironment{proof-idea}{\noindent{\bf Proof Idea}\hspace*{1em}}{\qed\bigskip}
\newenvironment{proof-of-lemma}[1]{\noindent{\bf Proof of Lemma #1}\hspace*{1em}}{\qed\bigskip}
\newenvironment{proof-attempt}{\noindent{\bf Proof Attempt}\hspace*{1em}}{\qed\bigskip}

\newcommand*\rot{\rotatebox{90}}

\newcommand\pa{\ensuremath{\boldsymbol{s}}} % protected attributes
\newcommand\out{\ensuremath{\boldsymbol{o}}} % application output

\newcommand\PA{\ensuremath{\boldsymbol{S}}} % protected attributes rv
\newcommand\UA{\ensuremath{\boldsymbol{X}}} % user attributes rv
\newcommand\EA{\ensuremath{\boldsymbol{E}}} % explanatory attributes rv
\newcommand\OUT{\ensuremath{\boldsymbol{O}}} % application output rv

\newcommand\mycode{\em}
\newcommand\investigation{{\mycode Investigation}\xspace}
\newcommand\testing{{\mycode Testing}\xspace}

\newcommand\discovery{{\mycode Discovery}\xspace}
\newcommand\errorprofiling{{\mycode Error Profiling}\xspace}
\newcommand\datasource{{\mycode DataSource}\xspace}

\newcommand\train{{\mycode train}\xspace}
\newcommand\test{{\mycode test}\xspace}
\newcommand\report{{\mycode report}\xspace}

\newcommand\Dtrain{{\ensuremath{D_{\text{train}}}}}
\newcommand\Dtest{{\ensuremath{D_{\text{test}}}}}

\usepackage{amsmath}
\usepackage{enumitem}
\usepackage{booktabs}
\usepackage[utf8]{inputenc}
\usepackage{authblk}
\usepackage{hyperref}
\usepackage[ruled,vlined]{algorithm2e}
\usepackage{fancyvrb}
\usepackage{cprotect}
\usepackage{array}
\usepackage{cite}
\usepackage{multirow}
\usepackage{etoolbox}
\usepackage{tabularx}
\usepackage[dvipsnames]{xcolor}

\def\sysname{FairTest\xspace}
\def\sec{\xspace\S}

\begin{document}

\date{}  % Don't want the date printed.

\title{
\sysname: Discovering Unwarranted Associations in Data-Driven Applications}

% [RG] I applied the following order: Florian first, followed by everyone else
% in alphabetical order.  I would have included Vaggelis second even if his
% last name was Zzz.  But I think the rest of us should probably be in alpha
% order, since no one has really stood out as the main non-student leader.
  \author[1]{Florian Tramèr}
  \author[2]{Vaggelis Atlidakis}
  \author[2]{Roxana Geambasu}
  \author[2]{Daniel Hsu}
  \author[1]{\\Jean-Pierre Hubaux}
  \author[4]{Mathias Humbert}
  \author[3]{Ari Juels}
  \author[1]{Huang Lin}
  \affil[1]{{\em EPFL}}
  \affil[2]{{\em Columbia University}}
  \affil[3]{{\em Cornell Tech}}
  \affil[4]{{\em Saarland University}}

\maketitle

\thispagestyle{plain}
\pagestyle{plain}

% ------------------ %
\begin{abstract}

In a world where traditional notions of privacy are increasingly challenged
by the myriad companies that collect and analyze our data, 
it is important that decision-making entities are held accountable
for unfair treatments arising from irresponsible data usage.
Unfortunately, a lack of appropriate
methodologies and tools means that even identifying unfair or 
discriminatory effects can be a challenge in practice.
%we must rely on
%those companies' responsible use of our data to ensure our fair 
%treatment.

We introduce the \emph{unwarranted associations (UA) framework}, 
a principled methodology for the discovery of
unfair, discriminatory, or 
offensive user treatment in data-driven applications.
The UA framework unifies and rationalizes a number of prior attempts at formalizing algorithmic fairness.
It uniquely combines multiple investigative primitives and fairness metrics with broad applicability, 
granular exploration of unfair treatment in user subgroups,
and incorporation of natural notions of utility that may account for observed disparities.

We instantiate the UA framework in {\em \sysname}, the first comprehensive tool that helps
developers check data-driven
applications for unfair user treatment.
It enables scalable and statistically rigorous investigation of
 associations between application outcomes (such as prices or
premiums) and sensitive user attributes (such as race or gender). 
% \sysname supports three types of investigations: \discovery of unwarranted
%associations, \testing of suspected associations, and \errorprofiling of predictive algorithms.
%In conjunction with these investigations, 
Furthermore, \sysname provides {\em debugging
capabilities} that let programmers rule out potential confounders for observed
unfair effects.

We report on use of \sysname to investigate and in some cases address
disparate impact, offensive labeling, and uneven rates of
algorithmic error in four data-driven applications.
As examples, our results reveal subtle biases against older populations
in the distribution of error in a predictive health application and
offensive racial labeling in an image tagger.

\end{abstract}

\section{Introduction}
\label{sec:introduction}

Today's applications collect and mine vast quantities of
personal information. Such data can boost applications' utility by personalizing content
and recommendations, increase business revenue via targeted product
placement, and improve a wide range of socially beneficial services, such as healthcare,
disaster response, and crime prevention.

The collection and use of such data raise two important
challenges.
First, massive data collection is perceived by many as a major threat to
traditional notions of individual privacy.
Second, the use of personal data for algorithmic decision-making can have
unintended and harmful consequences, such as unfair or discriminatory
treatment of users.

In this paper, we deal with the latter challenge. Despite the
personal and societal benefits of today's data-driven world, we argue that
companies that collect and use our data have a responsibility 
to ensure equitable user treatment. Indeed, European and U.S. regulators, as well as various policy and legal scholars, have recently called for 
increased \emph{algorithmic accountability}, and in particular for decision-making tools to be audited and ``tested for fairness''~\cite{algosAccountable, algosAccountableLaw}.

There have been many recent reports of unfair or discriminatory effects in
data-driven applications, mostly qualified as unintended consequences of data heuristics or overlooked bugs.
For example, Google's image tagger was found to associate racially offensive
labels with images of black people~\cite{googleGorilla}; the developers called
the situation a bug and promised to remedy it as soon as possible.
In another case~\cite{staples}, \emph{Wall Street Journal} investigators showed
that Staples' online pricing algorithm discriminated against lower-income
people.
They referred to the situation as an ``unintended consequence'' of Staples's
seemingly rational decision to adjust online prices based on user proximity
to competitors' stores. This led to higher prices for
low-income customers, who generally live farther from these stores.

Staples' intentions aside, it is evidently difficult for programmers to
foresee all the subtle implications and risks of data-driven heuristics.
Moreover, these risks will only increase as data is passed through increasingly
complex machine learning (ML) algorithms whose associations and inferences
may be impossible to anticipate.

We argue that such algorithmic biases are new kinds of {\em bugs}, specific to
modern, data-driven applications, that programmers should proactively
check for, debug, and fix with the same rigor as they apply to other
security and privacy bugs.
Such bugs can offend and even harm users, and cause programmers
and businesses embarrassment, mistrust, and potentially loss of revenue.
They may also be symptoms of a malfunction of a
data-driven algorithm, such as a ML algorithm exhibiting poor
accuracy for minority groups that are underrepresented in its training
set~\cite{hardt14}. 

We refer to such bugs generically as {\em unwarranted associations}.
Understanding and identifying unwarranted associations is an important step towards
holding automated decision-making entities \emph{accountable} for unfair practices,
thus also providing incentive for the adoption of corrective measures~\cite{algosAccountable, algosAccountableLaw, blackBoxSociety, datta2016algorithmic}. 

\paragraph{The Unwarranted Associations Framework.}
In order to reason about this new class of bugs, and propose preliminary means to fixing them, we first need an appropriate definitional framework in which to express them.

Substantial prior work exists on algorithmic
fairness~\cite{pedreschi08, ruggieri10,ruggieri10a,luong11,kamiran09,kamiran10,calders10,zliobaite11, hajian13, feldman14,
kamishima12, dwork12, zemel13}. 
The general approach is to characterize as ``fairness bugs''
any sufficiently strong statistical dependency, under a particular metric, between
an algorithm's outputs (\eg prices or labels) and protected user groups (\eg those defined by race, gender, or income level).
While the overall approach is sound, and partially underlies our own definitional framework, we find previously proposed definitions 
and metrics for fairness to be systematically lacking in at least one dimension of interest:
(1) Applicability to a broad range of settings, applications or data types, with rigorous statistical assessments; %For instance, the majority of prior works introduce metrics that only apply to  restricted settings in which both the algorithm's outputs and the considered protected groups are binary.
(2) Efficient and scalable assessment of disparate effects on a granular level, such as to discover biases affecting specific user subpopulations; and
(3) Inclusion of natural explanatory factors (\eg application or user ``utility'') that account for
apparent unfair effects.
%Most prior works consider \emph{any} strong enough statistical dependency as an indication of unfair treatment, thus disregarding important nuances ingrained in social and legal notions of fairness.
%and (4) Rigorous and computationally efficient discovery and statistical assessment of unfair effects.

To unify and rationalize the substantial list of prior work, and extend it to broader settings, we introduce the \emph{unwarranted associations (UA) framework}.
Informally, we define an {\em unwarranted association}, also called an {\em association  bug}, as any strong association 
between the outputs of an algorithm and features defining a protected user group, where the association arises in a meaningful subset of users and has no explanatory factors. (We give more details below.)

The UA framework proposes a principled methodology for testing for---and possibly debugging---unwarranted associations in a series of five steps:
(1) Data collection, and identification of user features and algorithmic outputs of interest; (2) Integration of \emph{explanatory factors}, typically special user or application requirements that justify ostensible unwarranted associations; (3) Mapping of the data and application of interest to an appropriate \emph{statistical metric} for assessing the strength of unfair associations;  (4) Testing for association bugs over semantically meaningful \emph{user subpopulations}, while ensuring \emph{statistical validity} and \emph{interpretability} of results; and finally
(5) All or part of steps 1-4 may be repeated for \emph{debugging purposes}, \eg through the addition of extra features or explanatory factors.

In this framework, we identify three core investigative primitives that permit broad  testing and debugging of association bugs in data-driven applications:
(1) {\em Discovery} of association bugs, with limited a priori
knowledge of what bugs to expect;
(2) {\em Testing} for suspected association bugs (\eg higher
prices or denied loans); 
and (3) {\em Error profiling} of an ML algorithm over a user
population, i.e., identifying subpopulations with disparately high error rates.

The UA framework specifically addresses three pressing needs noted by Kroll et al.~in their recent survey article on accountable algorithms~\cite{algosAccountableLaw}: (1) ``Expanding the repertoire of definitions of
group fairness that can be usefully applied in practice''; (2) ``providing better exploratory and explanatory tools for comparing different notions of fairness'' and (3) enabling ``system designers to have a set of rules, standards, or best practices that explain what notions of fairness are best used in specific real-world applications.''~

%The UA framework specifically addresses the growing need, noted by Kroll et al., for 
%``system designers to have a set of rules, standards, or best practices  that explain what notions of fairness are best used in specific real-world applications.''~\cite{algosAccountableLaw}

\paragraph{The \sysname Testing Toolkit.}
We instantiate our UA framework in {\em \sysname}, the first system that helps developers
test for and to some extent debug unwarranted associations in
data-driven applications.
Designed for ease of use, \sysname aggregates significant
unwarranted associations in an easily interpretable {\em bug report}, 
filtered and ranked by association strength, while accounting for known
explanatory factors.
%\sysname identifies both weak associations that affect large populations and
%strong associations that affect smaller subpopulations. 
%For example, our simulation of Staples' pricing scheme fed with data from the
%U.S. census reveals that while some disparate impact on low-income
%populations arises throughout the U.S., certain areas, such
%as New York state, exhibit stronger discriminatory effects.

\sysname supports the three core investigative primitives identified in the UA framework, and integrates a canonical set of carefully chosen association metrics, each suitable for particular applications and data types. In conjunction with these testing primitives, \sysname offers {\em debugging
primitives} that let a developer rule out potential confounders for observed
unfair effects. Our experience with four data-driven applications suggests that these primitives are effective at revealing and diagnosing subtle association bugs, such as skewed error against older patients in a predictive healthcare system and offensive racial labeling in an image tagger.

To efficiently identify semantically meaningful subpopulations affected by association bugs (step 4 in the UA framework), \sysname uses a novel technique we call {\em association-guided tree construction}. 
Inspired by decision-tree classifiers, our algorithm recursively splits
the user space into smaller subsets so as to maximize some metric of association between algorithm outputs and protected user attributes.
Each step yields subpopulations of decreasing size and
increasingly strong disparate effects.
We experimentally show that our tree-based algorithm finds
highly affected subpopulations by searching through up to $8$ times fewer candidates than previous ``unguided'' approaches~\cite{ruggieri10, ruggieri10a, luong11}.

\paragraph{Contributions.}
Our work offers the following contributions:
% to the space of algorithmic fairness, the closest domain to our work:

\begin{enumerate}
\item {\em The  UA framework}, and its principled methodology for discovering and analyzing association bugs in data-driven applications.
Our framework uniquely combines: Multiple primitives and metrics with broad applicability, explanatory factors, and fine-grained testing over user subpopulations, with rigorous statistical assessments, even for multiple adaptively chosen experiments.

\item {\em The  association-guided tree construction algorithm}, for efficiently finding meaningful and  interpretable user subsets strongly affected by algorithmic bias.

\item {\em The  design, implementation, and evaluation of \sysname},
a system that instantiates the UA framework, and 
the first testing and debugging tool for unwarranted associations.
We will release its source code on publication.

\item {\em Thorough experimentation with \sysname} across six applications 
and datasets, revealing the widespread occurrence of association
bugs, as well as \sysname's effectiveness in detecting them.
\end{enumerate}

% !TEX root = paper.tex
\section{Preliminaries}
\label{sec:motivation}

Developers routinely test and debug their programs for
functionality, security, and privacy bugs. 
Many supporting tools exist exist for these purposes.
Unfortunately, unwarranted associations constitute a newly emerging class of bugs in modern,
data-driven applications, and no good testing tools for them currently exist.
%These are unintended discriminatory effects which we generically call {\em unwarranted associations}.
%Lack of appropriate definitions and tools is an important obstacle towards
%increased accountability for unfair decision-making processes.
%We believe that as programmers take increasing liberties in collecting
%and mining our data, they become responsible for ensuring not only
%the data's security in face of malicious attack but also our fair and
%moral treatment when using the data optimize or improve their processes.
Our aim is to demonstrate the importance of proactively
testing for unwarranted associations in data-driven applications, 
and to provide appropriate definitional foundations and 
tools for investigating them.

\subsection{Motivating Examples}
\label{sec:motivation:use-cases}

%Typical examples of unwarranted associations in the related literature
%focus on high-stakes processes where differential treatment
%or impact is punishable by law, \eg hiring, providing credit, or
%offering housing. While such sensitive applications indeed require
%close inspection, we argue that {\em any} application that
%ingests and processes user data deserves scrutiny for association bugs.
We present three examples that underscore the diverse contexts
in which unwarranted associations can arise, and illustrate the testing
capabilities needed to detect them:

\begin{itemize}
\item{\em Google Photos.} Google's recently released Photos
  application includes an ML-based image tagging system.
  Users found that
  Photos produced offensive labels, tagging black people in
  photos as ``gorillas''. Google promptly
  apologized for the bug, saying that ``This is 100\% not OK,''
  and promised to fix it~\cite{googleGorilla}.

\item{\em Staples' differential pricing scheme.} The office retailer
  Staples implemented a seemingly rational differential
  pricing scheme for online purchases: users located
  within about 20 miles of a rival store (\eg
  Office Depot) were often offered discounted
  prices. The \emph{Wall Street Journal} (WSJ) found
  that the pricing scheme had a negative disparate impact on
  low-income customers, with results varying widely between states. WSJ called
  the situation an ``unintended side-effect''~\cite{staples}.

\item{\em Healthcare prediction.}  Based on 
real-world data and a winning approach from the
  Heritage Health Competition~\cite{healthprize}, we built a model
  using past healthcare claims to predict a user's number
  of hospital visits in a year.
  Using \sysname, we found that while the model has high accuracy
  overall, its errors are unevenly concentrated
  on elderly users, especially in subpopulations with certain
  pre-existing conditions.
  An insurance company that uses this algorithm to tune insurance
  premiums might involuntarily discriminate against these elderly
  patients.
\end{itemize}

These examples show the wide variety of unwarranted associations,
and the importance of proactively testing for and remedying them before
they can harm users or embarrass companies.
These settings also illustrate specific, significant advances needed to address such concerns.
Google's Photos case shows that unwarranted associations can be difficult
to anticipate, for algorithms with large output spaces, and therefore that 
we need sound methodologies and tools for {\bf \em Discovery}
of such bugs.
Staples' mishap illustrates the need, given a suspected
discriminatory effect, for methods that enable rigorous {\bf \em Testing} of
its presence, extent, and impact, across different meaningful user subpopulations.
Finally, the healthcare prediction example shows that rather than yielding
outcomes that are discriminatory in their content,
an application can provide disparate levels of utility to certain subpopulations,
such as uneven error rates. This scenario calls for principled {\bf \em Error
Profiling} of ML algorithms. More generally, it highlights the necessity to 
account for utility, when assessing unwarranted associations. 

%Our goal will be to propose a definitional framework
%and comprehensive system design, that capture these types of investigations, that we deem
%as valuable for finding and diagnosing unwarranted associations.

\subsection{Candidate Approaches and Related Work}
\label{sec:motivation:candidate-approaches}

A number of intuitive approaches may appear to allow discovery or prevention of association bugs. We review potential solutions
and highlight some common limitations.

\paragraph{Information Removal and Tracking.}
It is well known that explicit omission of protected
user attributes (e.g., gender, race) from an application's inputs is insufficient to avoid discrimination along these axes.
Indeed, subtle associations between
protected attributes and other program
inputs (e.g., location) may result in indirect biases.
For example, Staples' algorithm did not explicitly ingest information
about the socio-economic status of its customers; yet location surfaced
as an unanticipated {\em proxy} for this sensitive information.
For the same reason, traditional information-flow
tracking~\cite{sen14} is also insufficient to detect proxy-based
discriminatory effects.

\paragraph{Algorithmic Fairness.}
The literature on {\em algorithmic fairness} has proposed valuable
 techniques for both {\em preventing} and {\em detecting} unfairness
in data-driven applications.
However, our review of them reveals 
a fragmented set of definitions and concepts lacking unifying definitional foundations, and little work on
development and evaluation of usable tools for data-driven programmers.
We have carefully studied 13 representative works on algorithmic fairness:~\cite{kamiran09,kamiran10,calders10,
kamishima12,dwork12,zemel13,hajian13,pedreschi08,
ruggieri10,ruggieri10a,feldman14,zliobaite11,luong11}.
We observe four deficiencies in the field's current state:

\smallskip $\bullet$ {\em No generic, broadly-applicable criteria or metrics:}
The range of fairness criteria proposed in the literature is
large and fragmented.
Most works~\cite{ruggieri10, ruggieri10a, kamiran09,kamiran10,calders10,luong11, zliobaite11, hajian13, feldman14} use ratio and difference metrics, introduced by Pedreschi et al.~\cite{pedreschi08}. Other criteria include
a-protection~\cite{ruggieri10, ruggieri10a, hajian13},
e-fairness~\cite{feldman14}, statistical parity~\cite{dwork12,
zemel13} and mutual information~\cite{kamishima12}.
All of these criteria measure some form of association of {\em program
outputs} (\eg prices) on {\em
protected user attributes} (\eg race, gender).
Our review suggests that each of these metrics is relevant to specific situations and data types, yet none is universally applicable.
Proposed mechanisms typically apply only to specific chosen metrics, and thus cover a limited range of cases. For instance, most prior works only consider binary protected features and outputs~\cite{pedreschi08,
kamiran09,kamiran10,calders10, ruggieri10, ruggieri10a, luong11,
zliobaite11, hajian13, feldman14}.

Our UA framework unifies this fragmented landscape, by proposing a principled mapping from scenarios and data types of interest to appropriate statistical metrics.

%A line of work initiated by Dwork et al.~\cite{dwork12, zemel13} has
%further considered fairness from an \emph{individual's} point of
%view, essentially aiming at building classifiers that \emph{treat similar
%people similarly}. A fundamental issue with this approach is that fairness is
%defined with respect to a socially agreed-upon \emph{similarity
%metric} for users in a given context~\cite{dwork12}.
%Constructing this metric is non-trivial, especially using
%only information about individuals in a small sample, as
%opposed to \emph{every individual from the population}.
%It is also non-trivial to test proposed notion of fairness on
%a sample~\cite{jha2013testing,dixit2013testing}.

\smallskip $\bullet$  {\em Limited consideration of fine-grained discrimination contexts:}
Most prior works~\cite{kamiran09, kamiran10, calders10, zliobaite11, kamishima12, feldman14}
consider algorithmic fairness
solely at full user population level.
Yet, effects in a population
may differ from, and even contradict, those exhibited in
smaller subsets, an effect known as Simpson's paradox~\cite{simpson51}.
The 1973 Berkeley admissions~\cite{bickel75} are a famous example:
admission rates seemed to disfavor women, yet
individual departments showed either no bias or a reverse bias.
Some works~\cite{ruggieri10, ruggieri10a, luong11} recognized the need to proactively search for biases in user subsets, but do so through
exhaustive enumeration of meaningful population subsets, leading to a number of
contexts that is exponential in the feature space~\cite{ruggieri10,ruggieri10a}
or linear in the user space~\cite{luong11}.
%These works further fail to account for multiple-hypothesis testing for the many contexts they
%consider, raising concerns about the statistical validity of their results.

Fine-grained bias discovery is a core component of the UA framework, and we propose a novel algorithm, called association-guided tree construction, to efficiently search for user subpopulations exhibiting  strong unfair effects.

\smallskip $\bullet$  {\em Disregard for utility and plausible explanations for observed biases:}
Biases observed in a population may disappear when accounting for specific differences between groups of individuals.
In many situations, apparent discrimination can be explained by inherent differences in \emph{utility}, such as naturally opposing inclinations between protected groups, or fundamental application requirements that favor certain user demographics.
Some prior works do account for genuine application requirements~\cite{ruggieri10, ruggieri10a, luong11, dwork12, zliobaite11}. Yet all without exception explicitly define fairness as a difference in \emph{outcome proportions} between protected groups, thus rejecting the possibility that users may receive varying levels of utility from the same algorithmic outcomes. 

For instance, when profiling the error of an ML algorithm (one of our motivating examples), it is implicitly understood that an algorithm can be unfair, even if it provides all of its users with the same outcome, if this outcome happens to be the ``correct'' one for only a fraction of them.

\smallskip $\bullet$  {\em Limited experimentation:}
Prior works describe limited
experimental results, usually over two or fewer datasets~\cite{calders10, kamiran09, kamiran10, ruggieri10, ruggieri10a, zliobaite11, kamishima12, dwork12, hajian13}, and omit system design issues such as scalability and usability.
We believe that extensive experience with many applications and datasets is
crucial for developing a robust and flexible system, such as \sysname, that can address real-world algorithmic fairness issues. 

\paragraph{Web Transparency.}
Our work also relates to the field of {\it web
transparency}~\cite{Hannak:2013uk,
%Xing:2014ws,
Hannak:2014do,Vissers:2014uo,
Barford:2014ug,BinLiu:2013jn,
%Book:2015ue,
adFisher,Lecuyer:2014wj,LSSCGH15-sunlight, datta2016algorithmic}.
While some works touch on discrimination
and fairness (e.g.,~\cite{Hannak:2014do, Vissers:2014uo, adFisher, datta2016algorithmic}), their setting is different: These works rely on {\em controlled, randomized experiments} that {\em probe} a service
with different inputs (generally fake user profiles) and observe the effects on outputs,
so as to quantify Web services' use of personal data to target,
personalize, and tune prices.
Detection of unwarranted associations, as in \sysname, requires making
inferences from application behavior on {\em real user profiles}, which may
contain attribute correlations that do not
manifest in fabricated user profiles. For instance, a controlled experiment
of Staples' algorithm (\ie with fabricated profiles of low-income and high-income users)
would not uncover associations between prices and location, unless correlations
between location and income are explicitly accounted for.

\paragraph{Debugging of Machine Learning Applications.}
A number of methods exist in ML to understand data features
and modeling errors~\cite{ng2011advice,ng2013error,hastie2009elements,kabra2015understanding}.
These mechanisms are insufficient to identify and investigate unwarranted 
associations:
A challenge remains the systematic inspection of many combinations of
features that may define meaningful subpopulations.
Existing tools for deriving subpopulations (\eg
clustering algorithms~\cite{hastie2009elements}) are not guided by
measures that define discriminatory effects, and may
hence miss important effects.

\subsection{Goals and Assumptions}
\label{sec:motivation:assumptions}

\paragraph{Our Goals.}
In defining unwarranted associations and the UA framework (see \S\ref{sec:uaframework}),
our goal is not only to unify and rationalize the various definitions and metrics introduced in prior works,
but also to extend algorithmic accountability and fairness to a broader range of applications. For instance, by including the Discovery and Error Profiling primitives, consideration of utility, and support for adaptive debugging, we aim to cover scenarios that have received little attention in the literature.
In designing \sysname, our goal is to instantiate our proposed methodology in an efficient, broadly applicable, and easy-to-use system. \sysname should help developers test for---and to some extent debug---the types of bugs or unintended effects motivated in \S\ref{sec:motivation:use-cases}.

\paragraph{Assumptions and Limitations.}
The UA framework and \sysname are intended for  
use by honest developers, willing to find
unwarranted associations in their applications, \eg for accountability purposes.
We do not consider detection of unfair effects in 
applications that intentionally seek to conceal them.
Applying our methods in the context of \emph{adversarial machine learning} (\eg to
test for maliciously planted biases in training data)
is an interesting topic for future work.

When debugging uncovered association bugs, we restrict ourselves
to the identification of \emph{confounding} factors that might explain (and maybe justify) 
a bug's presence (\eg in the Berkeley
admissions~\cite{bickel75}, the department to which a person applied to was
a confounder). 
While this is valuable, confounders are not the only possible
``explanation'' for association bugs.
Other potential causes, such as insufficient training data for an ML algorithm, are out of scope.

While our framework and \sysname offer a methodology and tools
for association bug {\em detection} and
to some extent {\em debugging}, support for {\em 
remediation} is not explicitly covered.
Traditional testing tools (\eg for functionality bugs)
rarely prescribe fixes for the bugs they reveal;
developers use them to find and understand bugs, and then develop their
own fixes, which can range from trivial changes to major application
restructuring. 
In practice, with \sysname---as with other testing tools---fixes often become apparent once a developer understands a bug.
In \S\ref{sec:investigations} we show specific cases where debugging
yields clear remediation paths.

A key assumption in our framework is that a developer has at her disposal
a set of protected attributes (\eg gender or race) for her users.
In some cases, this information may be available
from user profiles.
Alternatively, public datasets (\eg census data) can
be leveraged to test for unwarranted associations on attributes that a
developer lacks (see\sec\ref{sec:report-example}).
Finally, one could imagine a deployment where a trusted auditor
(\eg the EFF) collects protected attributes from a user
population and runs \sysname on developers' behalf.
%In this case, the auditor is trusted by programmers and users.

\section{The UA Framework}
\label{sec:uaframework}

We now give a more detailed definition of unwarranted associations, inspired by legal and statistical practices. We then define a generic methodology for the discovery of such bugs in data-driven applications, and show how the investigative primitives motivated in \S\ref{sec:motivation:use-cases} fit into this framework.

\subsection{Unwarranted Associations}
\label{sec:unwarranted-associations}

Unwarranted associations are, as noted above, a new type of bug that
encompasses a broad class of unfair effects of data-driven
applications. 
To be able to test for such a bug, we need to define it.
In doing so, our goal is {\em not} to give a new mathematical specification of algorithmic discrimination; many characterizations are possible and useful, as we show.
Instead, we propose a definition flexible enough to encompass and extend a number of existing definitions and metrics, yet underpin a principled framework for identifying and ultimately resolving unwarranted associations.

\smallskip
We define an {\em unwarranted association} as any \textbf{statistically significant
association}, in a semantically meaningful \textbf{user subpopulation}, between a \textbf{protected attribute} and an
\textbf{algorithmic output}, where the association has no accompanying \textbf{explanatory factor}. 
\smallskip

\iffalse
We define an unwarranted association as {\em any \textbf{statistically significant
association}, in a semantically meaningful \textbf{user subset}, between a \textbf{protected attribute} (such as gender or race) and an
\textbf{algorithm outcome} or a quantity derived thereof (such as a price, a hiring decision, or an error rate)
that \textbf{cannot be explained} by any factor that is deemed as acceptable in the given context (\eg a natural
user inclination toward some class of products, an application requirement, etc.).}
\fi

This definition is broadly applicable, and rooted in legal discrimination practice.  For example, in U.S.
law, a ratio of 4/5 for hiring rates of two groups is generally considered discriminatory, but lower effects may
 qualify if statistically significant (and higher effects may be ignored if statistically insignificant)~\cite{ugesp78}.
Furthermore, the notion of explanatory factors reflects, e.g., the legal notion of
business necessity, which stipulates that
differential treatment of two groups may be acceptable if that difference can be shown to arise as
a consequence of fundamental business needs. Whether factors that account for observed effects are
deemed as acceptable or not remains foremost a matter of context and policy, which we do not attempt to
objectify here. 
%Indeed, while FairTest may take into account user-defined confounding variables in order to
%explain apparent discriminatory effects, we make no claims about the moral or legal validity of these factors.

Statistical association is a broad notion: it encompasses any
relationship between two measured quantities that renders them statistically
dependent~\cite{statsDico}.
This captures direct (causal) dependencies, indirect dependencies via proxies,
and relationships between any types of variables.
For each variable type, there are specific metrics known in the statistics
literature that can be applied to measure the degree of dependency.
For each metric, there are also known statistical tests to quantify
the statistical significance of an association.

In the following, we propose a principled methodology for the
discovery and analysis of association bugs, which formally specifies the various components that make up the above definition
(\ie protected features and outputs, statistical associations, user subsets, explanatory factors).

%To ensure \sysname's wide applicability to many applications and
%investigation types, we explicitly leave the particular metric to be used
%unspecified in our notion of unwarranted associations, as well as in
%\sysname's core design.
%As shown in \S\ref{sec:association-metrics}, \sysname can be instantiated
%using any of a variety of association metrics found in the literature, that
%appears best suited for a given context.

\subsection{The Methodology}
\label{sec:methodology}

Our methodology consists of five generic steps for discovering and investigating unwarranted associations. We detail these steps below, and highlight design considerations for the \sysname system that implements this methodology.

\paragraph{1) Data Collection and Pre-Processing.}
We consider applications that take inputs on users, such
as location or age, and return user-specific outputs.
To check an application for unwarranted associations, one begins by collecting
a dataset containing attributes from application
users, along with the outputs (or some derived quantity) for those users.

The output to be tested for associations is denoted $\OUT$.
The choice of $\OUT$ may reflect notions of \emph{user utility}, a point we expand on below. 
To ensure broad applicability, we require that 
our methodology and \sysname remain agnostic to the \emph{semantics}
of the quantity $\OUT$ to be investigated. 
Following~\cite{pedreschi08, ruggieri10, ruggieri10a}, we consider three types of user attributes:

\begin{enumerate}
\item {\em Protected attributes}, $\PA$, are discrimination-sensitive features 
(\eg race or gender) on which to look for associations with $\OUT$. What constitutes a protected attribute is foremost a matter of context, policy and law~\cite{ugesp78}. 
\item {\em Contextual attributes}, $\UA$, are dimensions along which the user population can be split into
semantically meaningful subpopulations (\eg location, profession, etc.), in which to search for associations $\OUT$ and $\PA$.
These include user attributes that are knowingly used in an
application and may also include protected attributes.
\item {\em Explanatory attributes}, $\EA$, are user properties on which
it is deemed acceptable to differentiate, even if that leads to
apparent discrimination on protected attributes. Explanatory attributes play a critical role in expressing fundamental application requirements and in the debugging of previously found
association bugs.
\end{enumerate}

\paragraph{2) Integration of Explanatory Factors.}
Associations between protected features and application outputs
may not always be indications of a bug, but rather natural consequences of utility requirements.
As noted above, the outputs of interest $\OUT$ can (and should) be chosen so as to reflect
the utility perceived by application users. On the one hand, all users may perceive similar utility
from receiving a discount price or a positive hiring decision. On the other, a health prediction may only be useful to a user if it is accurate.

Some associations may also be deemed acceptable or necessary if they are induced by fundamental application requirements, \eg a hiring decision based on a person's qualifications, which could  be an inadvertent proxy for gender or age. Such explanatory factors can be expressed through explanatory attributes $\EA$ on which the statistical association between $\PA$ and $\OUT$ can be conditioned.

\newcolumntype{L}[1]{>{\raggedright\let\newline\\\arraybackslash\hspace{0pt}}p{#1}}
\begin{table}[t]
  \centering
  \begin{tabular}{@{} m{4.6cm} m{4.85cm} m {6.2cm} @{}}
  	{\bf Metric} & {\bf Description} & {\bf When to Use} \\
  	\toprule
	Binary Ratio \& Difference\newline\cite{pedreschi08,kamiran09,kamiran10,calders10,ruggieri10,ruggieri10a,luong11,zliobaite11,hajian13,feldman14} & compare probability of an output
	for 2 groups & binary $\PA,\OUT$; often for {\em Testing}\\
	\midrule
	Mutual Information (MI)~\cite{kamishima12} & dependence measure for 2 discrete variables &categorical $\PA,\OUT$; often for {\em Testing}\\
	\midrule
	Pearson Correlation (CORR) & measures linear dependence between 2 scalar variables &scalar $\PA,\OUT$; often for {\em Error Profiling} \\
	\midrule
	Regression & for labeled outputs, measure association for each label &high dimensional $\OUT$;
	always for {\em Discovery}\\
	\bottomrule
  \end{tabular}
  \caption{ {\bf Canonical Association Metrics for the UA Framework}.}
  \label{tab:metrics}
\end{table}

\paragraph{3) Selection of Appropriate Metrics.} 
Given a chosen quantity $\OUT$ to be tested for associations against a protected attribute $\PA$,
the next step is to select a statistical metric that is appropriate for measuring the strength or extent of
the association between $\OUT$ and $\PA$ in a given context.

As noted in \S\ref{sec:motivation:candidate-approaches}, prior work has identified a number
of metrics appropriate for measuring associations between certain types of quantities $\OUT$ and $\PA$.
Table~\ref{tab:metrics} shows a set of five canonical metrics that we selected
after careful examination of the relevant literature, two of which are new to the algorithmic fairness space.
Organized based on $\PA$ and $\OUT$ types, the metrics can be
classified in three categories:

\smallskip
$\bullet$ {\em Frequency Distribution Metrics:}
    Association between nominal $\PA$ and $\OUT$ 
    (with few possible values) can be represented as a
    \emph{contingency table} that displays the variables' frequency distribution.
    Prior work used such tables to define \emph{ratio and difference metrics}
    for binary variables~\cite{pedreschi08,kamiran09,kamiran10,calders10,ruggieri10,ruggieri10a,luong11,zliobaite11,hajian13,feldman14}.
For $\OUT=\{ \out_1,
\out_2 \}$ and $\PA=\{ \pa_1, \pa_2 \}$, the \emph{ratio
metric} is $\Pr(\out_1|\pa_1) / \Pr(\out_1|\pa_2) - 1$, and the
\emph{difference metric} is $\Pr(\out_1|\pa_1) -
\Pr(\out_1|\pa_2)$.    
    These are difficult to extend to non-binary
    protected attributes~\cite{dwork12}, such as race.
    In general cases, we measure association with \emph{mutual
    information} (MI), given by
  $
  \sum_{\out,\pa}
  \Pr(\out,\pa)
  \ln
  \big(
    \frac{
      \Pr(\out,\pa)
    }{
      \Pr(\out)
      \Pr(\pa)
    }
  \big)$
    To compare effects across
    multiple associations, we use a normalized version of MI (NMI), obtained by dividing the measure by the minimum of the Shannon entropies of $\PA$ and $\OUT$, 

\smallskip$\bullet$ {\em Correlation:}
    Measuring dependence of scalar variables (\eg with MI~\cite{poczos2011nonparametric})
     is hard, so it is common to
    consider specialized relationships for such variables.
    {\em Pearson's correlation} measures the strength of linear
    associations between $\OUT$ and $\PA$, which may exist even for non-linearly related variables~\cite{rodgers1988thirteen}.
    %These measures are often robust and broadly
    %interpretable~\cite{rodgers1988thirteen}.
    Note that a finding of zero correlation does \emph{not} imply
    independence.
    However, as our aim is not to verify independence, and
    as we value interpretability, Pearson's
    correlation is a natural fit.
    We can use correlation, for instance, to test for associations between a patient's age
    and the error incurred by a healthcare predictor.

\smallskip$\bullet$ {\em Regression:}
    High-dimensional output spaces occur in many use-cases, such
    as for applications that assign tags or labels to users, where it
    is not known \emph{a priori} which specific tag/label to test for
    associations (see the Discovery investigation examples in
    \S\ref{sec:motivation:use-cases}).
    For these, we introduce a metric based on \emph{regression}.
    We model the relationship between the
     protected attribute $\PA$ and a large number of
    dependent output {labels} $\OUT$ with a regression model (logistic or linear).
    This yields a \emph{regression coefficient} for each label, with
    which we can estimate that label's association with $\PA$.
    
    We detail our approach for binary $\PA$ and applications
	that output $t$ labels, each in $\mathcal{L} = \{l_1, \dotsc, l_d\}$ 
	($\OUT$ takes values in $\mathcal{L}^t$).
	Let $b_1, \dotsc, b_d$ be indicator variables
	for each label ($b_i = 1 \Leftrightarrow l_i \in \OUT$).
	We model the distribution of $\PA$
	given $\OUT$ by $\Pr[\PA=1\ |\ b_1, \dotsc, b_d] =
	\operatorname{logistic}(\beta_0 + \sum_{i=1}^d \beta_i b_i)$, where
	the $\beta_i$ are regression coefficients that measure the association between the labels and the protected feature~\cite{logistic}.
	%This approach extends to multi-valued or continuous $\PA$ by replacing
	%logistic regression with multinomial or linear regression.
	
\smallskip
If an {\em explanatory attribute} $\EA$ is identified, 
these metrics can be made to measure the \emph{conditional
association} of $\PA$ and $\OUT$, given $\EA$.
Conditional metrics quantify the association between $\PA$ and
$\OUT$ that remains after controlling for $\EA$.

\paragraph{4) Granular and Statistically Rigorous Testing across User Subpopulations.}
As argued in \S\ref{sec:motivation:candidate-approaches}, testing for associations solely in the full user population is insufficient, as strong unfair effects may manifest only in specific user subsets. 
The core testing step in our framework consists in finding strong associations between protected features and outputs, in specific subpopulations of application users. 

How should a ``subpopulation'' be defined? If we simply consider arbitrary subsets of users, it is trivial to group users so as to obtain maximal associations between some protected feature and output (\eg for associations between gender and income, the subset of users formed by high-income women and low-income men exhibits a maximal association).
Instead, to ensure that subpopulations are semantically meaningful and easily interpretable, we restrict our attention to subsets of the global population obtained by partitioning users based on the value of a few contextual features $\UA$ (\eg the subset of users with `\texttt{Job} $=$ \texttt{Researcher}' and `\texttt{Age} $\leq$ \texttt{50}'). Such ``association contexts'' enable an analysis of algorithmic fairness among users that share a number of similar characteristics.

Efficiently searching for subpopulations with high associations is non-trivial: Any user subset defined by features $\UA$ may exhibit very different effects, a consequence of Simpson's paradox. Where prior work has considered unguided, exhaustive search of potential contexts~\cite{ruggieri10, ruggieri10a, luong11}, \sysname uses a novel algorithm, called \emph{association-guided tree construction}, to successively split users into smaller subsets with increasingly strong associations (\S\ref{sec:context-generation}).
We experimentally compare our algorithm to prior work in \S\ref{sec:performance-benchmarks}.

When assessing associations in multiple subpopulations, care has to be taken to ensure the results remain statistically valid. To this end, \sysname applies well-known techniques, such as hypothesis testing, cross-validation and rigorously correcting for multiple statistical inferences (\S\ref{sec:design:stats}).

\paragraph{5) Adaptive Debugging.}
Following the previous steps, a developer may find evidence of association bugs affecting certain application users. 
The question remains on how such bugs may be \emph{debugged} or \emph{fixed}. Maybe the presence of the bug itself is sufficient to alert a developer to a specific issue in her application, allowing her to correct the flaw. Alternatively, a discovered bug could hint to an overlooked explanatory feature that accounts for its presence. In any case, a new investigation would have to be run (following the above methodology) to verify that the bug was effectively removed.
As these debugging investigations are inherently informed by results from previous investigations, ensuring statistical validity of discoveries requires extra care.
We discuss the system design implications for \sysname, as well as our preliminary solution to this problem in \S\ref{sec:dataset-management}.

\subsection{Three Core Investigation Primitives}
\label{sec:primitives}

We show how the core investigative primitives identified in \S\ref{sec:motivation:use-cases} map onto the UA framework methodology:

\paragraph{Testing.} This is the simplest investigative primitive in the UA framework. It tests for suspected associations (under a suitable metric such as MI or correlation) between application outputs $\OUT$ and protected user features $\PA$, possibly conditioned on an appropriate explanatory feature $\EA$.

\paragraph{Discovery.} This primitive applies to scenarios
in which a developer may not know the specific association bugs to test for.
An example are applications that tag users with labels from a large space. 
None of the metrics proposed in prior work are appropriate for
extracting biases  between protected features and a large set of candidate outputs.
We thus propose a new regression-based metric that efficiently estimates the strength of the
association between protected attributes and each output label and allows the identification of the most highly associated labels.
In this sense,
\discovery requires little \emph{a priori} knowledge of
what could constitute an association bug or what subpopulations it might affect.

\paragraph{Error Profiling.} This is a form of Testing where the quantity tested for associations is not the algorithm's output, but rather each user's individual error. This primitive illustrates how the quantity $\OUT$ can be defined in terms of the \emph{utility} that users perceive from application outputs: a predictive output is presumably only useful to a user if it is reasonably correct.

% !TEX root = paper.tex

% Figure source is on Google Docs (feel free to edit it): https://docs.google.com/presentation/d/1nzH3nwWhaY1RIOCQ2YaTkPQkxhHr1IZzZgcrNICgdY8/edit#slide=id.p
% Latex version of the algo is in: fairtest-algo.tex.
\begin{figure*}[t]
  \includegraphics[width=\linewidth]{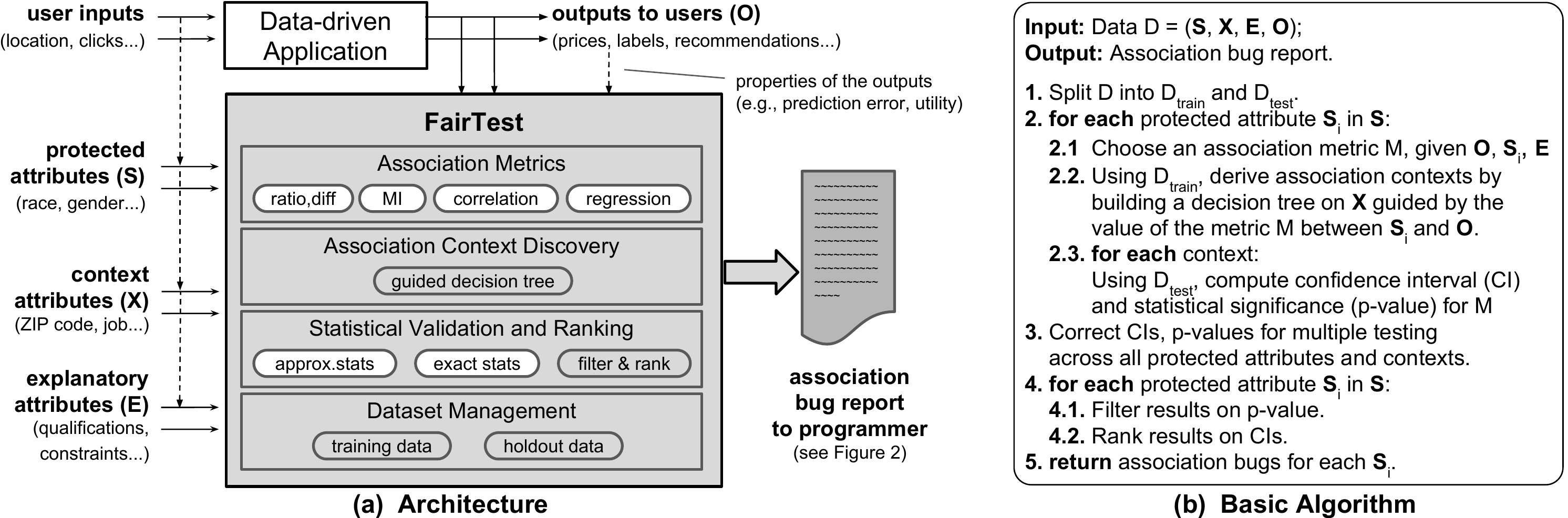}
  \caption{
    {\bf \sysname Architecture.}
     (a) Grey boxes denote \sysname components. Rounded boxes denote
     specific mechanisms.  White rounded boxes denote extensibility
     points; transparent rounded boxes denote core, generic mechanisms.
      (b) \sysname's basic algorithm, realizing the UA framework methodology. $\PA$, $\UA$, $\EA$ denote protected, context, and explanatory attributes, respectively; $\OUT$ denotes the output quantity of interest.
  }
 
  \label{fig:design}
\end{figure*}

\section{The FairTest Design}
\label{sec:design}

We describe the design of \sysname, a system that realizes the UA framework methodology
and incorporates the three core investigation primitives described previously.
\F\ref{fig:design} shows the \sysname architecture.
To test her application for unwarranted associations, a developer supplies
\sysname with a dataset consisting of attributes from application
users, along with the outputs (or derived quantities) for those users.
This corresponds to the first step of the UA methodology.
%Additionally, the choice of outputs $\OUT$ and explanatory attributes $\EA$ may reflect
%natural utility considerations, as previously described in \S\ref{sec:methodology}, 

\sysname analyzes this data and returns an {\em association report} that lists strong associations that \sysname has
found between specified protected attributes $\PA$ and the outputs $\OUT$.
The programmer inspects the report and determines which reported associations
are real bugs that require fixing and which are admissible effects in the
context of her company's policies.
After giving a concrete example of association report, we detail \sysname's architecture and illustrate how
it realizes the remaining steps of the UA methodology.

\begin{figure*}[t]
\fbox{
\begin{minipage}{0.6\textwidth}
\ttfamily

\textbf{Report of associations of O=Price on S\(\sb{i}\)=Income:}\\
\textbf{Association metric:\ norm.\ mutual information (NMI).}\\[0.5em]
\textbf{Global Population of size 494,436}\\
p-value = 3.34e-10 ; NMI = [0.0001, 0.0005]\\[0.5em]
\begin{tabular}{@{}l | r | r | r@{}}
Price & Income <\$50K & Income >=\$50K &    Total\\
\hline
High &   \textbf{15301 (6\%)}&    \textbf{13867 (6\%)}& 29168 \ (6\%)\\
Low  &  234167(94\%)&   231101(94\%)&465268 (94\%)\\
\hline
Total&  249468(50\%)&   244968(50\%)&494436(100\%)
\end{tabular}
\vspace{1em}

\textbf{1.\ Subpopulation of size 23,532}\\
\textbf{Context = State:\ CA, Race:\ White}\\
p-value = 2.31e-24 ; NMI = [0.0051, 0.0203]\\[0.5em]
\begin{tabular}{@{}l | r | r | r@{}}
Price & Income <\$50K & Income >=\$50K &    Total\\
\hline
High &     \textbf{606 (8\%)}&      \textbf{691 (4\%)}&  1297 \ (6\%)\\
Low  &    7116(92\%)&    15119(96\%)& 22235 (94\%)\\
\hline
Total&    7722(33\%)&    15810(67\%)& 23532(100\%)
\end{tabular}
\vspace{1em}

\textbf{2.\ Subpopulation of size 2,198}\\
\textbf{Context = State:\ NY, Race:\ Black, Gender:\ Male}\\
 p-value = 7.72e-05 ; NMI = [0.0040, 0.0975]\\[0.5em]
\begin{tabular}{@{}l | r | r | r@{}}
Price & Income <\$50K & Income >=\$50K &    Total\\
\hline
High &      \textbf{52 (4\%)}&        \textbf{8 (1\%)}&    60 \ (3\%)\\
Low  &    1201(96\%)&      937(99\%)&  2138 (97\%)\\
\hline
Total&    1253(57\%)&      945(43\%)&  2198(100\%)
\end{tabular}
\vspace{1em}

\textbf{... more entries (sorted by descending NMI) ...}
\end{minipage}
}

\caption{ {\bf Sample Association Bug Report.}
% Full population and two highly affected subpopulations for a
Testing investigation of disparate impact in Staples pricing simulation.
}
\label{fig:sample-bug-report}
\end{figure*}

\subsection{Association Report Example}
\label{sec:report-example}

Suppose that Staples' developers wished to inspect their
pricing scheme's impact on users before deploying it
(e.g., to avoid bad publicity or for accountability purposes).
To do so, they could use U.S. census statistics~\cite{census} to emulate users
with realistic demographics visiting their website from various locations.
They would run their location-based pricing scheme for those users and use
\sysname's {\em Testing investigation} to test for disparate impact on race
or income.

% XXXXXXXXXXX: BEFORE SHRINKING/ALTERING THIS SECTION, PLEASE NOTE:
%
% 1. The goal is to show-case an end-to-end scenario of how a programmer
%   can use the tool to find something interesting, and how he can act on the
%   finding (e.g., disable or add noise to the algorithm for problematic
%   subpopulations).
% 2. Make people understand the structure of our bug reports.
% 3. One place where programmers might get their protected attributes in
%    certain cases (US census).
% 3. Convince people that it's important to look at subpopulations (which is
%    why I say explicitly that the global population doesn't reveal an
%    interesting .

We ran an investigation on a simulated pricing scheme akin
to Staples', which gives discounts to users located within 20 miles of
competing OfficeDepot stores.
\F\ref{fig:sample-bug-report} shows part of \sysname's bug report,
generated by testing for suspected differential pricing based
on income.
The report lists statistically significant associations discovered
between protected attribute `\texttt{income}' and output `\texttt{price}' in
various subpopulations.
Association strength is measured with {\em normalized mutual
information} (NMI), one of the canonical measures of statistical dependence implemented in \sysname.

The report shows three populations: the global population is first, followed
by the two subpopulations exhibiting the strongest disparities (highest NMI).
Subpopulations are defined by user attributes: white people in California
(first subpopulation) and black men in New York (second subpopulation).
For each (sub)population, \sysname reports varied statistical information:
a p-value (a measure of statistical significance, with a value below $5\%$
generally considered statistically significant), a confidence interval for the
NMI metric, and a {\em contingency table} that summarizes the frequency
distribution of the outputs over the (sub)population.

Results can be read from these tables as follows. For the first subpopulation:
among California's white population ($23{,}532$ people in our test set),
$7{,}722$ (or $33\%$) have an income below $\$50$K. Out of these $7{,}222$ users, 
$606$ (or $8\%$) got the high price and the rest ($92\%$) got a discount.
% Similarly for the other entries.

The report can be interpreted as follows:
``At global U.S. population level, the disparate impact of the pricing
scheme against lower-income users is nearly zero (NMI is close to
zero for the global population; low-income and high-income
users receive high prices in similar proportion, 6\%).
Yet, disparities are much stronger among white people in California
(first subpopulation), where 8\% of lower-income people get
higher prices vs.\ only 4\% of higher-income people.
Strong disparities also exist for black men in New York
(second subpopulation): 4\% of lower-income black men get higher prices
vs.\ 1\% for higher-income black men.''
As remediation, a programmer may decide to alter her pricing scheme, e.g.,
to disable price tuning in affected regions,
%\footnote{Incidentally, WSJ reports that Staples did not appear to
%apply their pricing scheme to users from Alaska~\cite{staples}.
%Our own race-based report (Appendix~\ref{sec:staples-app-details})
%shows that applying the scheme would have resulted in severe %discrimination
%against Native Americans across that state.}
% [FT] can't find a source for this
or to take into account public statistics for various areas when
deciding prices for people in those areas.
\subsection{Architectural Components}

\F\ref{fig:design}(a) illustrates \sysname's inputs and architectural components, which map onto the 
UA framework methodology introduced previously.
\sysname's core (\sysname box in \F\ref{fig:design}) consists of four
architectural components ({\em Association Metrics}, {\em Association Context
Discovery}, {\em Statistical Validation and Ranking}, and {\em Dataset
Management}), detailed in forthcoming sections.
At a high level, \sysname generates reports using
the algorithm in \F\ref{fig:design}(b), which makes use of the aforementioned components
as follows.

\begin{itemize}
\item
The dataset $D=\{(\PA, \UA, \EA, \OUT)\}$ is first split into
a {\em training set}, $D_{\text{train}}$ and a {\em testing set},
$D_{\text{test}}$ (Step 1).

\item 
The {\em Association Metric} module, which incorporates the association
metrics from Table~\ref{tab:metrics}, chooses a metric that will be used to quantify the dependency of
the output on a protected attribute (Step 2.1).

\item 
For each protected attribute $\PA_i$ in $\PA$, the {\em Association Context
Discovery} module uses $D_{\text{train}}$ to split the user population
into meaningful subpopulations based on contextual attributes, $\UA$ (Step 2.2).
Its goal is to maximize the association between $\PA_i$ and $\OUT$ in the
discovered subpopulations.
For that, it uses our new association-guided decision tree
construction algorithm (\S\ref{sec:context-generation}).

\item
For each discovered subpopulation, the {\em Statistical Validation and Ranking}
module assesses the bug's validity on $D_{\text{test}}$ by applying an
appropriate test statistic (Steps 2.3 and 3).
% The choice of test statistic depends on the previously selected association
% metric.
To prioritize developers' efforts, the module also filters and ranks
 bugs according to the strength of the measured associations (Steps 4-5).
% (\S\ref{sec:design:stats}).

\item The {\em Dataset Management} module manages the testing sets
to guarantee statistical validity of \sysname's results across multiple
investigations that follow Steps 1-5.
\end{itemize}

\subsubsection{Association Metrics}
\label{sec:association-metrics}

This module incorporates a canonical (yet extensible) set of statistical association metrics (Table~\ref{tab:metrics}) and selects the appropriate
metric for each investigation, as prescribed by the third step of the UA methodology.
Unless specified by the user, a metric is chosen automatically based on the data types of protected attributes
$\PA$ and outputs $\OUT$, the presence of explanatory attributes $\EA$, and the
investigation type.

% !TEX root = paper.tex
\subsubsection{Association Context Discovery}
\label{sec:context-generation}

\setlength{\algomargin}{0em}
\begin{algorithm}[t]
\DontPrintSemicolon
\SetAlCapHSkip{0em}
%\SetKwProg{Fn}{Function}{}{}
\SetKwFunction{findContexts}{findContexts}
\SetKwFunction{metric}{Metric}
\SetKwBlock{Function}{Function \findContexts{$D=\{\PA, \UA, \EA, \OUT\}$, $\mathcal{P}=\emptyset$}}{end}
\SetKwData{Left}{left}\SetKwData{This}{this}\SetKwData{Up}{up}
\SetKwInOut{Params}{Params}

\Params{\texttt{MIN\_SIZE} \tcp*[r]{Minimum size of a context}\\
\texttt{MAX\_DEPTH} \tcp*[r]{Maximum tree depth}\\
\metric \tcp*[r]{Association metric}}
\vspace{3pt}
\Function{
\vspace{2pt}
Create a subpopulation defined by predicates $\mathcal{P}$ \\
\lIf{$\ |D| < \texttt{MIN\_SIZE}$\ or\ $|\mathcal{P}| \geq \texttt{MAX\_DEPTH}\ $}{
\Return
}
\For{$\UA_i \in \UA$}{
$\mathbb{D} = \{D_1, D_2, \dots\} \gets $ partition of $D$ based on the value of $\UA_i$ \\
\uIf{\(\exists\ D_k \in \mathbb{D}: \metric{$D_k$} > \metric{$D$}\)}{
\tcp*[l]{Average association for this split}
Score$_i$ \(\gets\sum_{D_k \in \mathbb{D}}\metric{$D_k$}/|\mathbb{D}|\) 
}
\lElse{\hspace{1pt}Score$_i \gets 0$}
}
\If{\ $\forall i:$ Score$_i \leq \metric(D)$\ }{
\Return \tcp*[r]{No partition yields a higher association}
}
$\UA_{\text{best}}, \mathbb{D}_{\text{best}} \gets $ partition with highest score \\
\vspace{1pt}
\For{$D_k \in \mathbb{D}_{\text{best}}$}{
\vspace{2pt}
 $V \gets $ values taken by $\UA_{\text{best}}$ in $D_k$ \\
 \findContexts{$D_k$, $\mathcal{P} \cup \{\UA_{\text{best}} \in V\}$}
}
}

\caption{
    {\bf Association-Guided Decision-Tree Construction}
    We build increasingly specific contexts (subpopulations) with increasingly stronger associations, by recursively splitting the data $D$ upon the user attribute $\UA_i$ that maximizes the average association over derived contexts $\mathbb{D}$. Contexts are defined by predicates $\mathcal{P}$ over attributes $\UA$ (tree paths).
    %We control over-fitting by specifying a minimal size for a tree leaf (and thus of a context) and a maximal tree depth (equivalently the maximal number of predicates defining a context).
  }

\label{algo:findcontexts}
\end{algorithm}

This module efficiently
``zooms into'' a user population, to discover subpopulations (a.k.a. contexts)
strongly affected by association bugs.
In prior work, finding such bugs has required uninformed exhaustive enumeration
of user subsets, leading to
a number of contexts either exponential in the feature space~\cite{ruggieri10, ruggieri10a} or
linear in the user space~\cite{luong11}.
These methods raise two concerns: (1) They require making a large number
of statistical inferences, thus providing only weak guarantees on the false discovery rate.
(2) They sacrifice the ability to discover small subpopulations (\eg a few hundred users), 
even if these exhibit the highest associations.

To efficiently identify strong hidden associations (the fourth step in the UA methodology), 
we develop a novel partitioning scheme, called {\em guided decision-tree construction}.
In contrast to prior work, our method generates only a \emph{constant}
number of contexts, while actively searching for smaller, highly affected
populations.
\Alg\ref{algo:findcontexts} shows our algorithm.
Similarly to how decision-tree learning 
greedily optimizes some measure of
target \emph{homogeneity} (\eg Gini impurity)~\cite{c45}, our
algorithm actively maximizes some \emph{association metric}
between protected attributes and outputs.

%Let $\PA_i \in \PA$ be a sensitive attribute under investigation.
The algorithm works by selecting a splitting rule, based on an attribute
%$\UA \cup \PA \setminus \{\PA_i\}$
$\UA_i \in \UA$, so as to split the dataset $D$ into
subsets $\mathbb{D} = \{D_1, D_2, \dots\}$ with highest average association between $\PA$ and
$\OUT$. For continuous $\UA_i$, we split the data into two subsets, based on some threshold; if $\UA_i$ is categorical, we split the data into one subset per value of $\UA_i$.
We only consider a split if it yields at least
one sub-context with a higher association than the one measured over
the current population. 
%To prioritize splits that lead to many
%potential association contexts, we assign each split a score defined
%as the average association strength over all derived sub-contexts.
%on the subsets are as high as possible (on average, where the
%subsets are weighted by their relative sizes).
We then recursively apply this process on each subset derived from the highest scoring split.
This approach:
(1) permits use of \emph{any} association metric;
(2) produces simply-defined and interpretable subpopulations;
and
(3) aggressively searches for subpopulations with strong associations using just
scalable/distributable computations~\cite{mllib}.

We additionally employ well-known techniques for preventing this tree
construction from overfitting the training data~\cite{c45},
such as bounding the tree's depth and pruning very small subpopulations
($<100$ members).

\subsubsection{Statistical Validation and Ranking}
\label{sec:design:stats}

After discovering contexts that exhibit potential association bugs,
we validate and rank them before reporting them to developers.
Validation is needed because the contexts were explicitly built so as to maximize associations over a
finite sample of users ($\Dtrain$).
We validate bugs on an independent test sample ($\Dtest$).

We use distinct notions of significance for bug validation and ranking.
For validation, we use statistical significance based on {\em hypothesis
testing}: a bug is significant if its manifestation in the test set is
unlikely under the ``null hypothesis'' (\ie the association between $\PA$ and
$\OUT$ is null).
This is quantified by the {\em p-value} for a test.
For ranking, we use {\em effect size}, \ie the actual value of the association
metric, estimated by means of a {\em confidence interval} (CI).
\sysname incorporates statistical methods for computing p-values and CIs for
all metrics in~\S\ref{sec:association-metrics}; for small samples, we
use generic {\em permutation tests}~\cite{ernst2004permutation} and
bootstraps~\cite{efron1979boostrap} instead of approximations.
We apply Holm-Bonferroni corrections~\cite{holm1979simple} to the p-values and
CIs to ensure their simultaneous validity.

A bug report is then generated as follows:
We filter out contexts with corrected p-values ${>}0.05$ (configurable).
We rank the remaining contexts by the lower bounds of their corrected effect-size CIs.
We only include a context (\eg white males in NY) if it exhibits a stronger
effect than the larger populations (\eg males in NY) that contain it.
Eventually, the report prioritizes debugging efforts by 
(1) listing only statistically significant associations
and (2) first displaying
the most strongly affected subpopulations.

\subsubsection{Dataset Management}
\label{sec:dataset-management}

% Our decision to add debugging support in \sysname raises subtle
% issues with significant (and quite unexpected) implications for system
% design.
To debug an association bug, developers may run multiple
investigations, each informed by previous ones:
A bug is detected in a first investigation, after which
a series of other analyses (\eg with explanatory
attributes) are run to narrow down its source.
Yet, while the test set $\Dtest$ is independent from hypotheses
formed in the first investigation, it is {\em not} independent from
hypotheses formed in subsequent adaptively-chosen ones. Such adaptive debugging is the challenge raised by the 
final step of the UA methodology.

%Because investigations use knowledge gained in previous steps, it is
%incorrect from a statistical perspective to validate them on the same test
%set as previous investigations.  %; it risks overfitting the testing set.

The system-design implications are significant: \sysname
cannot be a
stateless library; it must manage the dataset it is given across
multiple investigations.
We achieve this by having developers specify a budget
$B$ (number of adaptive investigations they plan to run) upfront, when they supply a
dataset; \sysname then splits the dataset into $B$ testing sets, each used for a single investigation.
Only $B$ investigations are allowed on the same dataset.
The \datasource abstraction in
\F\ref{fig:api} implements this functionality.
More efficient approaches have been proposed~\cite{dwork2015reusable}, but they
all lead to similar systems implications, restrictions, and interfaces.

\subsection{Testing and Debugging Primitives}

\begin{figure}[t]
\begin{lstlisting}[xleftmargin=4pt,framexleftmargin=2pt,
		language=python,
        keywordstyle=\color{blue},
        commentstyle=\color{OliveGreen},
        rulecolor=\color{black},
        framesep=0pt]
class DataSource(D, budget) `\ \ ` # hold out 1 testset per budget unit
\end{lstlisting}
\vspace{-5pt}
\begin{lstlisting}[xleftmargin=4pt,framexleftmargin=2pt,
		language=python,
        keywordstyle=\color{blue},
        commentstyle=\color{OliveGreen},
        rulecolor=\color{black},
        framesep=0pt]
class Investigation(DataSource,S,X,E,O,M={}) `\ \ ` # base investigation class
  # M stands for association metrics
class Testing(DS,S,X,E,O,M) `\ \ ` # Investigation subclass for testing
class Discovery(DS,S,X,E,O,M,top_k) `\ \ ` # subclass for discovery
  # takes in number of outputs to consider in each context.
class ErrorProfiling(DS,S,X,E,O,M,groundTruth) `\ \ ` # Subclass for error profile
  #  takes ground truth for a predictive output
\end{lstlisting}
\vspace{-5pt}
\begin{lstlisting}[xleftmargin=4pt,framexleftmargin=2pt,language=python,
        keywordstyle=\color{blue},
        commentstyle=\color{OliveGreen},
        rulecolor=\color{black},
        framesep=0pt]]
train(Investigations,maxDepth=5,minLeafSize=100)
  # derive putative contexts for one/more investigations
test(Investigations,conf=0.95)  # test and correct associations
report(Investigations,conf=0.95,outDir)  # filter, rank, save reports
\end{lstlisting}
\vspace{-5pt}
\begin{lstlisting}[xleftmargin=4pt,framexleftmargin=2pt,language=python,
        keywordstyle=\color{blue},
        commentstyle=\color{OliveGreen},
        rulecolor=\color{black},
        framesep=0pt]]
class Metric `\ \ ` # abstract class for association metrics
Metric.computeStats(data, conf) # calculate p-value and CI
\end{lstlisting}

\caption{ {\bf\sysname API.}
Data holdout (1st), investigation types (2nd), methods to run investigations (3rd),
API to implement for metrics (4th).}
\label{fig:api}
\end{figure}

To test and debug unwarranted associations, \sysname provides the three core investigative primitives
defined in \S\ref{sec:primitives}.
\S\ref{sec:investigations} shows how these primitives can be combined together
to support quite powerful analyses of unwarranted associations, from
discovery to diagnosis and remediation.

%Below, we detail \sysname's API, that programmers use to launch investigations.
%We also provide further details on \sysname's debugging capabilities, centered
%around the use of \emph{explanatory features} (step 2 of the UA methodology).

\subsubsection{Investigation API}
\label{sec:api}

\F~\ref{fig:api} shows \sysname's API.
A core \investigation class is subclassed by three specific investigation
types.
To run these, a developer first collects user attributes
and application outputs.
This data is enclosed in a \datasource, which holds out test
sets for successive experiments (see\sec\ref{sec:dataset-management}).

\paragraph{Testing.} This investigation type was 
used in\sec\ref{sec:report-example} to test for disparate impact in
Staples' pricing scheme.
A developer provides a dataset $D=\{(\PA, \UA, \EA, \OUT)\}$, and \sysname first finds contexts
with potential associations between $\PA$ and $\OUT$ (possibly conditioned on $\EA$),
using the association-guided tree construction.
The developer calls the \train method to initiate context discovery,
optionally tuning the size and complexity of the resulting contexts.
She calls \test and \report to validate discovered bugs and
produce reports.

\paragraph{Discovery.} This lets developers
search for associations over a large number of output labels, using the regression metric from
\S\ref{sec:methodology} to efficiently estimate the strength of the
association between protected attributes and each output label.
We then select the \emph{top\_k} labels that exhibit the strongest associations,
and test each label individually using an appropriately chosen metric.
These regressions are executed at every step of the guided decision-tree
recursion. In the end, \discovery finds subpopulations in which 
a particular label is disproportionately associated with a protected group.

\paragraph{Error Profiling.} This searches for
populations strongly affected by algorithmic mistakes. This may help
improve both the algorithm's accuracy and fairness, as we show
for our healthcare predictor (\S\ref{sec:medical}).
\errorprofiling takes as inputs algorithm predictions and ground truth, and computes a suitable error metric to be tested for associations.

\subsubsection{Explanatory Attributes}
\label{sec:debugging}

\begin{figure}[t]
\fbox{
\begin{minipage}{0.6\textwidth}
\ttfamily

\textbf{Report of associations of O=Admitted on S\(\sb{i}\)=Gender,}\\
\textbf{conditioned on explanatory attribute E=Department:}\\[0.5em]
\textbf{Global Population of size 2,213}\\
\textbf{p-value = 7.98e-01} ; COND-DIFF = [-0.0382, 0.1055]\\[0.5em]
\begin{tabular}{@{}l | r | r | r@{}}
Admitted &  Female & Male & Total\\
\hline
No      &615(68\%)& 680(52\%)&1295 (59\%)\\
Yes     &\textbf{295(32\%)}& \textbf{623(48\%)}& 918 (41\%)\\
\hline
Total   &910(41\%)&1303(59\%)&2213(100\%)\\
\end{tabular}
\vspace{	1em}

\textbf{* Department A:\ Population of size 490:}\\
%\begin{addmargin}[1.25em]{0em}
\hspace*{8pt}\textbf{p-value = 4.34e-03} ; DIFF = [0.0649, 0.3464] \\[0.5em]
\hspace*{8pt}\begin{tabular}{@{}l | r | r | r@{}}
Admitted &  Female & Male & Total\\
\hline
No      & 9(15\%)&161(37\%)&170 (35\%)\\
Yes     &51(85\%)&269(63\%)&320 (65\%)\\
\hline
Total   &60(12\%)&430(88\%)&490(100\%)\\
\end{tabular}
\vspace{1em}

%\end{addmargin}
\textbf{* Department B:\ Population of size 279:}\\
%\begin{addmargin}[1.25em]{0em}
\hspace*{8pt}\textbf{p-value = 1.00e+00} ; DIFF = [-0.4172, 0.3704] \\[0.5em]
\hspace*{8pt}\begin{tabular}{@{}l | r | r | r@{}}
Admitted &  Female & Male & Total\\
\hline
No      & 3(30\%)& 93(35\%)& 96 (34\%)\\
Yes     & 7(70\%)&176(65\%)&183 (66\%)\\
\hline
Total   &10 (4\%)&269(96\%)&279(100\%)\\
\end{tabular}
\vspace{1em}

%\end{addmargin}
\textbf{* ... Departments C-F, all with high p-values ...}

\end{minipage}
}

\caption{ {\bf Disparate Admission Rates in the Berkeley Dataset.}
%Results from an initial investigation, which reveals the apparent
%bias in the full population (top);
Shows a \testing investigation with explanatory attribute $\EA$ $=$ $\operatorname{Department}$.
COND-DIFF is the binary difference metric (DIFF), conditioned on $\EA$.}
\label{fig:berkeley-bug-report}
\end{figure}

Following the use-cases detailed in \S\ref{sec:methodology}, \sysname lets developers use explanatory attributes in two ways:\
(1) After an initial investigation that reveals apparent unfair effects, she may \emph{debug} these associations by specifying explanatory attributes that she believes are responsible (\ie confounders) for the observed behavior. \sysname then recomputes conditional association metrics over the \emph{same} contexts discovered in the first investigation.
(2) The developer can define user properties $\EA$ that are necessary for an
application. \sysname then explicitly avoids
deriving associations that are accounted for by these attributes, by
measuring the dependence of protected attributes $\PA$ and outputs $\OUT$ \emph{conditioned} on $\EA$.

To illustrate the first use-case of explanatory attributes, we
examine the Berkeley graduate admissions dataset, which
contains admission decisions and gender for 4,425
applicants~\cite{bickel75}.
As mentioned in \S\ref{sec:motivation:candidate-approaches}, this data exhibits a paradoxical effect: at full university level, admissions appear to disfavor women,
yet this bias is not reflected in any department.
We show how an analyst could use \sysname to measure gender-disparities in admission rates,
while allowing that each department may have different gender demographics and admission rates.

The analyst defines `\texttt{department}' as an \emph{explanatory attribute}, to instruct \sysname to look for associations only among applicants of a same department. She then runs a \testing investigation. The report (\F\ref{fig:berkeley-bug-report}) clearly highlights
the paradox: Over the full population, only 32\% of female applicants are admitted versus 48\% for males. Yet, the only department with a significant
 disparity in admission rates (department `A') actually favors
women! Incidentally, the difference in admission rates \emph{conditioned} on an applicant's department is not statistically significant (p-value of $0.798$).

\subsection{Summary}

With this design, \sysname realizes the UA framework methodology as follows:
{\bf 1)} By keeping our design aware of the different types of user features ($\PA, \UA, \EA$), yet agnostic to the semantics of these features and of the algorithm's outcomes, we ensure \sysname's compatibility with the data-collection stage proscribed by the UA framework.
{\bf 2)} By providing debugging primitives based on the use of explanatory features as confounders, \sysname instantiates the second step of the UA methodology.
{\bf 3)} By making our design extensible with multiple metrics, we
ensure \sysname's applicability to a broad range of applications and
scenarios.
{\bf 4)} \sysname's novel association-guided tree construction algorithm 
enables efficient, scalable, and statistically rigorous detection of 
strongly affected subpopulations.
{\bf 5)} By keeping state of the data used in successive debugging investigations,
\sysname provides basic support for adaptive data analysis.

\iffalse
and meets the requirements listed
in \S\ref{sec:requirements} as follows.
{\bf R1:} By making our design extensible with any type of metric, we
ensure \sysname's applicability to a broad range of applications and
investigation types.
{\bf R2:} By guiding subpopulation construction by the specific association
metric used in an investigation, we can efficiently and scalably detect
strongly affected subpopulations.
{\bf R3:} By providing both testing and debugging primitives, and managing
the dataset carefully, we ensure both the effectiveness and the statistical
correctness of investigations with \sysname.
{\bf R4:} By being agnostic to the particular semantics of the outputs $\OUT$,
\sysname allows for easy incorporation of notions of user utility.
The explanatory feature primitive lets programmers specify
factors that may justify algorithmic bias.
\fi

% !TEX root = paper.tex

\section{Evaluation}
\label{sec:evaluation}

We implemented \sysname in Python, to be used as a
standalone library or as a RESTful service.
As a library, \sysname is designed to integrate
with Pandas, SciPy's data analysis library.
Our service prototype enables continuous monitoring for association
bugs in production systems, \eg for accountability purposes.
Developers register investigations with the service and route
user attributes and outputs to it.
\sysname runs periodically and sends reports to
developers. The monitoring service can be continuously
replenished with new test data, thus supporting larger numbers of
investigations than the standalone library running on static datasets.

We focus our evaluation on the standalone library, the more mature of our two prototypes. 
We aim to address three questions:
{\bf (Q1)} Is \sysname effective at detecting association
bugs?
{\bf (Q2)} Is it fast enough to be practical? and
{\bf (Q3)} Is it useful to identify and to some extent debug
association bugs in a variety of applications?
We use seven workloads:

\begin{itemize}
\item {\em One tightly controlled microbenchmark}, which we use to evaluate
\sysname's bug detection abilities with a priori known ground truth for the
associations.

\item {\em Four data-driven applications} fed by public datasets:
(1) a simulator of Staples' pricing scheme (described in~\cite{staples})
fed by U.S. census data; (2) a predictive healthcare
app, based on a winning method and data from the Heritage Health Competition~\cite{healthprize};
and (3) an image tagger based on Caffe~\cite{caffe},
fed by ImageNet~\cite{imagenet}.
and (4) a movie recommender trained on MovieLens data~\cite{movielens}.

\item {\em Two social datasets} -- the Adult Census dataset~\cite{uci_repo} and the 1973 Berkeley
Admissions dataset~\cite{bickel75}.
% -- which have been used in prior algorithmic fairness work. 
%We use these to demonstrate the social analyst use case.

\end{itemize}

\noindent
Table~\ref{tab:applications} shows workload information: number
of users/attributes, investigations we ran, and metrics we used.
% We will detail these applications and datasets as needed for each experiment.
% We use our controlled microbenchmark to answer the first question, our
% real/realistic applications and datasets for the second, and a combination of
% them all for the third.

\subsection{Detection Effectiveness (Q1)}
\label{sec:accuracy-benchmarks}

\begin{figure}[t]
{
  \centering
    \includegraphics[width=0.7\columnwidth]
        {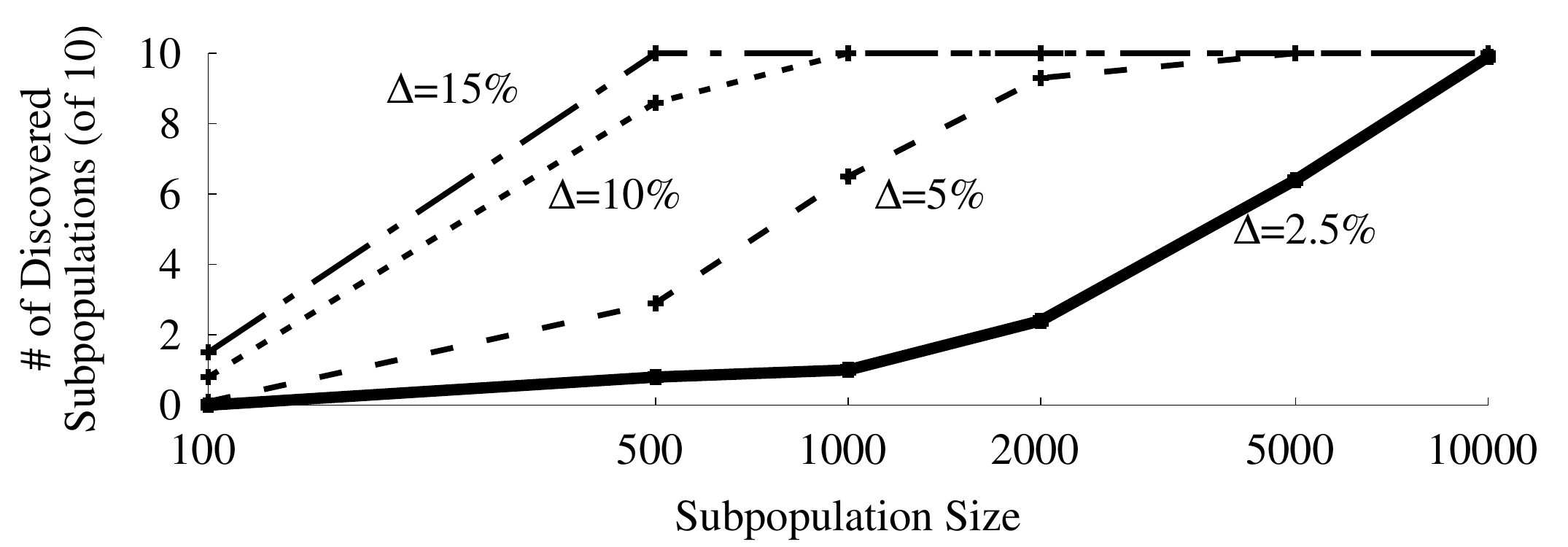}
    }
    \caption{{\bf \sysname Effectiveness with Affected
    Subpopulation Size and Effect Strength ($\Delta$).}
    Number of contexts discovered out of the ten
    we artificially inserted in 1M-user population.
    Average over 10 trials.}
    \label{fig:fairtest_recall}
\end{figure}

\paragraph{Microbenchmark.}
Inspired by the Staples case, we create a microbenchmark that lets us control
the strength and span of association bugs.
We use U.S.~Census~\cite{census} data for
gender, race, and income to generate $\approx1M$ synthetic users.
We start from a ``fair'' algorithm that randomly provides users
with $\{0,1\}$-output, independent of income.
We then plant disparities in certain subpopulations (determined
by location and race), so that income level (high or low) implies a difference
in output proportions of size $2\Delta$. E.g., for $\Delta=10\%$, we might give
output ``1'' to $60\%$ of high-income users and $40\%$ of low-income users, in the 
subpopulation of white users in California.
For various subpopulation sizes and effect sizes, we inject 10
randomly chosen discrimination contexts into our data and measure how
many are discovered by \sysname.

\F\ref{fig:fairtest_recall} shows \sysname's discovery rate as we increase
 population size and $\Delta$.
%Larger values of $\Delta$ mean stronger disparities.
\sysname reliably detects strong disparities that affect a few
hundred users, as well as effects as low as $2.5\%$ in larger contexts.
Low effects in small contexts often go undetected due to
limited statistical evidence.
% [RG] The sentence below is not useful.  We'd need a measurement of this.
%Note that if a programmer wishes to make a large number of \emph{adaptive} investigations, the size of each test-set is diminished in consequence (see \S\ref{sec:design:data-management}). Effects in small contexts are then more likely to go undetected.
In all cases, \sysname made zero false discoveries (finding a disparity that we did not introduce). Statistical testing lets us tightly control the false discovery rate: at a confidence level of $95\%$, we expect at most $5\%$ false discoveries.

\begin{table}[t]
  \centering
  \begin{tabular}{@{}l c r r c r r r r @{}}
  	&&&&& \multicolumn{4}{c}{{\bf Association Contexts}}\\
  	\cmidrule{6-9}
  	{\bf Application} & {\bf Investigations} & {\bf Users} & {\bf Attr.} & {\bf Metric(s)} & \rot{\bf Discovered} & \rot{\bf Validated} & \rot{\bf Reported} & \rot{\parbox{2cm}{\bf Smallest\\ Reported}}\\
  	\toprule
  	Microbenchmark & T & $988871$& $4$ & NMI & n/a & n/a & n/a & n/a\\
  	\midrule
	Staples Pricing & T & $988871$&$4$& NMI & $224$ & $100$ & $21$ & $211$\\
	%Health Prediction & EP & $86359$&$128$& NMI,CORR & $28$ & $24$ & $4$\\
	Predictive Healthcare & EP & $86359$&$128$& CORR & $33$ & $33$ & $2$ & $91$\\
	Image Tagger & D,T & $2648$&$1$& REG,DIFF & $1$ & $1$ & $1$ & $1324$\\
	Movie Recommender & T & $6040$&$3$ & CORR & $15$ & $10$ & $7$ & $511$\\
	\midrule
	Adult Census & T & $48842$ & $13$& NMI & $108$ & $57$ & $10$ & $104$\\
  	Berkeley Admission & T & $4425$&$2$& DIFF & $1$ & $0$ & $1$ & $2213$\\
	\bottomrule
  \end{tabular} 
  
  \caption{ {\bf Workloads.} {\rm
  Investigations: {\em Discovery} (D), {\em Testing} (T), {\em ErrorProfiling} (EP).
  Metrics: normalized mutual information (NMI), correlation (CORR), binary difference (DIFF), regression (REG). For each application, we report the number of potential association contexts found by \sysname's guided-tree construction, the number that were found to be statistically significant (p-value $<5\%$), and the number of reported bugs.}}
  \label{tab:applications}
\end{table}

\paragraph{Real-World Apps and Datasets.}
Table~\ref{tab:applications} reports the number of association contexts found
by \sysname in each application. We show the number of potential bugs found by
the guided decision-tree mechanism, the number of associations that are
statistically significant after correcting for multiple testing, and the
number of bugs reported to the developer (recall that we only report a context
if it exhibits a higher unfair effect than the larger subpopulations that
contain it).  The size of the smallest reported context is also shown 
(the largest reported context is the full test set).
We do not have ground truth for these workloads, but our experience
inspecting the reports (detailed in\sec\ref{sec:investigations}) suggests
that \sysname detects discrimination contexts of a variety of sizes,
which appear accurate and revelatory for an investigator.

Results for the predictive healthcare application are for an experiment
with a follow-up debugging investigation (see \S\ref{sec:medical}).
\sysname thus splits the dataset in
three: a train set and two test sets.
We found that \sysname would have reported the same bugs, had we
used all the data for a single investigation (\ie with no debugging).
We further analyzed the effect of the debugging budget $B$ on the number of discovered
bugs for the Staples application. For budgets $B$ of $2$ and $3$,
we discover $168$ and $125$ contexts, respectively; of these, we report
$15$ and $13$ contexts, respectively.
In both cases, the most affected subpopulation is the
same as the one found for a budget $B=1$. Thus, for this application,
\sysname can allow at least one or two follow-up analyses, while preserving
the main results reported to developers.

\iffalse
Our results (\F\ref{fig:fairtest_recall}) reveal both
strengths and weaknesses for \sysname:
Strong disparities that affect
at least a few hundred users, as well as disparities as low as $2.5\%$ over large contexts are persistently
found.
However, \sysname can not easily discover such low effects in small contexts,
due to the inherent limited statistical power in these cases.
(This also motivates the minimum subpopulation size that \sysname
uses in \train.)
\fi
%
%We note that the inability of \sysname to discover low effects in small sub-populations is mainly due to the limited \emph{power} of our statistical assessments for low sample sizes (in a sample of $500$, it isn't unlikely for a difference of $\pm2.5\%$ to appear purely by chance). As \sysname splits the available sample into training and testing sets, the sample size over which we validate our hypotheses is further reduced. For this reason, it makes little sense using \sysname to look for sub-populations smaller than a couple hundred users, unless we expect to find extremely high disparities.

\subsection{Performance (Q2)}
\label{sec:performance-benchmarks}

\begin{figure}[t]
{
  \centering
    \includegraphics[width=0.8\columnwidth]
        {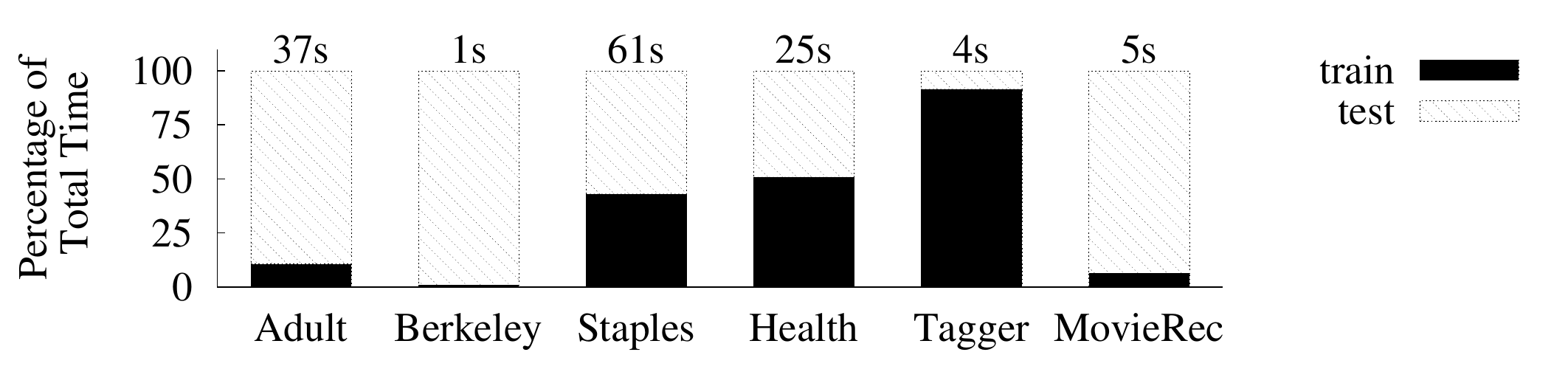}
    }
    \caption{{\bf \sysname Performance.}
     Total analysis time (labels above bars) broken down
     into training and testing times (bars).}
    \label{fig:timing_percentages}
\end{figure}

\paragraph{Timing.}
Although its building blocks (decision trees, statistical tests) admit
efficient and scalable implementations, our prototype does not incorporate all
available optimizations.
Still, \sysname appears fast enough for practical use.
\F\ref{fig:timing_percentages} shows the analysis time for
each application (top numbers), broken down into:
 (1) the time spent on training to form association hypotheses,
and (2) the time spent on testing and correcting these hypotheses.
On a commodity laptop (4-core Intel CPU @1.7GHz, 8GB RAM),
the total execution time ranges from 1-5 seconds for the smallest datasets
to 60 seconds for the largest (Staples, with 1M users).
For small datasets (Adult, Berkeley, Movies) we often use
bootstraps and permutation tests to compute CIs and p-values in
small contexts ($\leq$1000 users); these are expensive and subsume the training cost.
For datasets that yield larger contexts, we use faster, approximate methods, making the testing phase fast and the training phase
proportionally more expensive.
%For the Health dataset, training is particularly expensive due to many user features (128), most serving as contextual
%features in the decision tree's generation.
% Further performance benchmarks are in\sec\ref{sec:appendix:performance}.

\paragraph{Subpopulation Discovery.}
To discover subpopulations with high associations, Ruggieri et al.~\cite{ruggieri10, ruggieri10a}  use a simple data mining approach: They list all \emph{frequent itemsets} over contextual features, and consider each itemset as a potential discrimination context. As this enumeration of contexts is not informed by any association metric, it can be viewed as a brute-force analog to the guided tree mechanism used in \sysname.
Their approach suffers from two main caveats. First, it requires that all features be discretized. Second, it leads to an exponential increase in the number of contexts to consider, unless subpopulations of small size are neglected.

For a fair comparison between our guided-tree mechanism and the data mining approach, we analyze their performance when processing the \emph{same} number of contexts. We proceed as follows: We run \sysname, and record the number $N_{\rm train}$ of contexts examined in the training phase (\ie the number of metric computations in \Alg\ref{algo:findcontexts}) and the number $N_{\rm test}$ retained for testing. \sysname considers contexts with at least \texttt{MIN\_SIZE} users (\eg $100$ or $500$), and defined by at most \texttt{MAX\_DEPTH} features (\eg $5$).
For the data-mining approach, we list all itemsets with support at least \texttt{MIN\_SIZE} and at most \texttt{MAX\_DEPTH} items. We choose $N_{\rm train}$ such itemsets \emph{at random}, compute the association metric in those contexts over the training set, and retain the $N_{\rm test}$ contexts with highest effect for testing.
We compare effect-sizes of the associations found by \sysname, to the ones found with the ``brute-force'' approach, over the testing set. Limiting the unguided approach to only $N_{\rm train}$ random contexts may appear arbitrary. We thus further compare the two methods for larger numbers of considered contexts, by simply repeating the above process with fresh random choices. Both \sysname and the unguided approach then consider the same total number $\gg N_{\rm train}$ of potential contexts.

We use the Adult Census data, with protected feature ``Gender'', target ''Income'', and the binary difference metric. We add random permutations of the remaining features to obtain a dataset with either $15$ or $40$ contextual features (to see how the two methods scale with the number of features). The number of itemsets with at least $100$ users, and defined by at most $5$ features, grows from $84{,}000$ (for $15$ features) to over $200$ million for $40$ features.
For efficiency reasons, we thus limit our search to subpopulations of at least $500$ users, of which there are roughly $3$ million (for $40$ features).

\begin{figure}[t]
	{
	\centering
    \includegraphics[width=0.8\columnwidth]
        {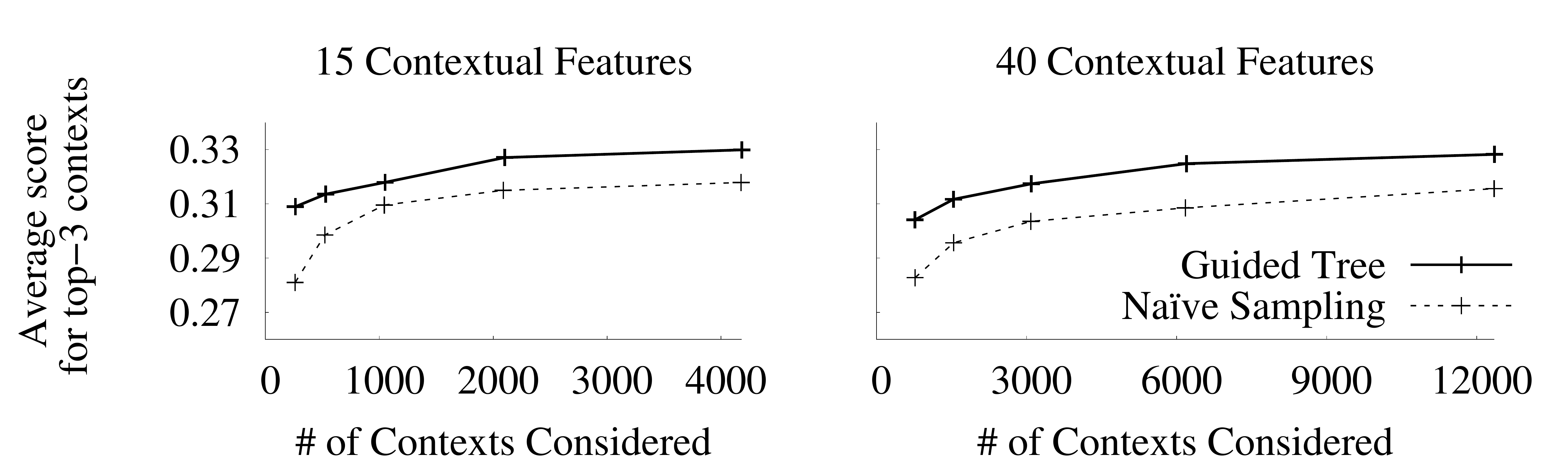}
    }
    \caption{ {\bf Comparison between \sysname's guided tree mechanism and naïve subpopulation sampling.} Shows the average of the $3$ highest associations scores uncovered in the testing phase, for an increasing number of candidates considered in the training phase.}
    \label{fig:naive}
\end{figure}

In \F\ref{fig:naive}, we plot the average association (as per the difference metric) for the $3$ most affected subpopulations discovered by \sysname, compared to those found by the unguided approach. The process is repeated $2,4,8$ and $16$ times, resulting in both methods observing more potential contexts (x-axis) and discovering increasingly strong associations (y-axis). In summary, \sysname finds similarly high associations as with the itemset mining approach, by considering $4\times$ to $8\times$ fewer potential contexts. Furthermore, it is always more optimal to simply re-run \sysname than it is to search for more potential contexts in an unguided fashion.

% !TEX root = paper.tex
% \vspace{-0.3cm}
\subsection{Investigation Experience (Q3)}
\label{sec:investigations}
% \vspace{-0.3cm}

% Q3: Is \sysname useful (and usable) for identifying and debugging
% association bugs in a variety of applications?

To assess \sysname's usefulness,
we searched for unwarranted associations in
all real-world applications in Table~\ref{tab:applications}.
The Staples scheme was described in \S\ref{sec:report-example}.
The others are detailed below.
Our experience shows that \sysname: (1) discovers meaningful bugs;
(2) it lets developers include acceptable explanations
for associations; and (3) it assists in debugging and even remedying discovered bugs.

More specifically, we find insightful association bugs in three of the four applications
 and showcase how to debug and remedy them in the case of the predictive
healthcare application. For the movie recommender application, we show how
apparent bugs discovered in a preliminary investigation can be explained by natural
notions of user utility.

\subsubsection{Predictive Healthcare}
\label{sec:medical}

Our predictive health application uses methods and data from winners of the Heritage Health Prize Competition~\cite{marketmakers, healthprize}.
The random-forest based algorithm uses past healthcare claims to predict a
user's number of hospital visits in the next year (predictions are for $\log(1+\text{number of visits})$).
The algorithm has low error overall (the average difference between the true and predicted number of visits is $0.42$), but we want to study the error's distribution among users.

Our study gives an end-to-end view of how \sysname can be
used to detect and debug unwarranted associations, but also obtain hints
for potential fixes.  (1) We first discover an association bug: the
application has much higher error rates for older than for younger users.
(2) We investigate the bug's source: the bias can be explained by lower
prediction confidence for older people.
(3) From there, we suggest potential fixes, such as only using high-confidence
predictions.
% Our study consists of two investigations (detection and debugging), which
% we perform adaptively, each on its own test set.
% We describe the specific process and results next.

\begin{figure}[t]
\fbox{
\begin{minipage}{0.7\textwidth}
\ttfamily
\centering

\parbox{\textwidth}{\textbf{Report of associations of O=Abs.\ Error on S\(\sb{i}\)=Age:}}\\
\parbox{\textwidth}{\textbf{Global Population of size 28,930}}\\
\parbox{\textwidth}{p-value = 3.30e-179 ; CORR = [0.2057, 0.2432]}

\vspace{0.5em}
\includegraphics[scale=0.5]{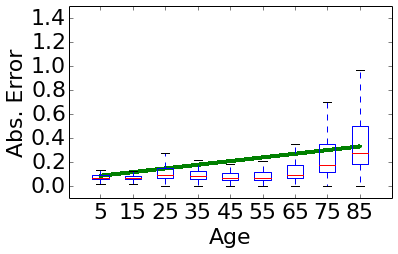}
\vspace{0.5em}

\parbox{\textwidth}{\textbf{1.\ Subpopulation of size 6,252}}\\
\parbox{\textwidth}{\textbf{Context = Urgent Care Treatments >= 1}}\\
\parbox{\textwidth}{p-value = 1.85e-141 ; CORR = [0.2724, 0.3492]}

\vspace{0.5em}
\includegraphics[scale=0.5]{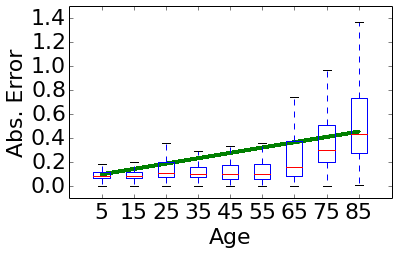}

\end{minipage}
}

\caption{ {\bf Error Profile for Health Predictions}. Shows the
global population and the subpopulation with highest effect (correlation). Plots display correlation between age and prediction error, for predictions of $\log(1+\text{number of visits})$.
For each age-decade, we display standard box plots (box from the 1st to 3rd quantile, line at median, whiskers at $1.5$ IQRs). The green line depicts the best linear fit over the data.\vspace{-0.5em}}
\label{fig:medical-bug-report}

\end{figure}

\begin{figure}[t]

\fbox{
\begin{minipage}{0.7\textwidth}
\ttfamily
\centering

\parbox{\textwidth}{\textbf{Report of associations of O=Abs.\ Error on S\(\sb{i}\)=Age,}}\\
\parbox{\textwidth}{\textbf{conditioned on explanatory attribute E=Confidence:}}\\[0.5em]

\parbox{\textwidth}{\textbf{Global Population of size 28,930}}\\
\parbox{\textwidth}{p-value = 1.26e-13 ; COND-CORR = [0.1050, 0.1597]}\\[0.5em]

\parbox{\textwidth}{\textbf{* Low Confidence:\ Population of size 14,481}}\\
\parbox{\textwidth}{\hspace*{8pt}p-value = 2.27e-128 ; CORR = [0.1722, 0.2259]}

\vspace{0.5em}
\includegraphics[scale=0.5]{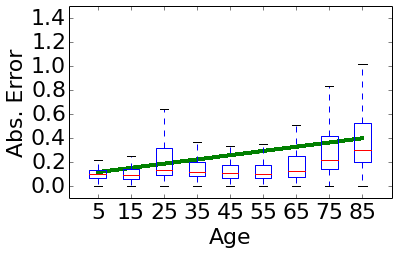}
\vspace{0.5em}

\parbox{\textwidth}{\textbf{* High Confidence:\ Population of size 14,449}}\\
\parbox{\textwidth}{\hspace*{8pt}p-value = 2.44e-13 ; CORR = [0.0377, 0.0934]}

\vspace{0.5em}
\includegraphics[scale=0.5]{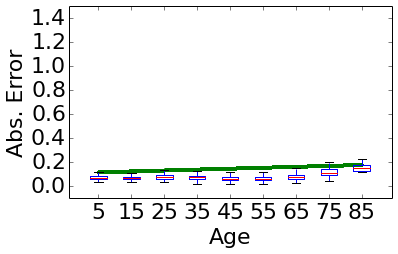}

\end{minipage}
}

\caption{ {\bf Error Profile for Health Predictions using prediction confidence as an explanatory attribute}. Shows correlations between prediction error and user age, broken down by prediction confidence.}
\label{fig:medical-bug-report-explained}
\end{figure}

\paragraph{Detection.}
We first use \sysname's \errorprofiling to examine associations between
the algorithm's prediction error and a user's age (scalar quantities, hence we use correlation).
The report (\F\ref{fig:medical-bug-report}) shows the error/age
correlations for
the full user population and one subpopulation with higher effect.
We visualize correlation with plots instead of contingency tables.
Globally, prediction error grows with age (correlation is positive and the data shows a clear positive linear trend).
% [FT] The line isn't exactly related to correlation
This effect is strongest for patients with prior urgent-care treatments.
In that context, the average error for patients of age 61-99 is $1.07$,
compared to $0.33$ for younger patients.

This finding is alarming, as such disparities could cause quantifiable
harms if, e.g., the algorithm is used to adjust insurance premiums (one of the competition's motivations~\cite{healthprize}).
Hence we wish to further investigate the {\em causes} of this
 accuracy loss for older patients, and get insights into how to fix this
fairness (and accuracy) bug.

\paragraph{Debugging.}
We use \sysname's debugging abilities (explanatory attributes) to
verify a plausible cause for the observed bias:
Higher errors for older users could be due to a higher \emph{variance} of the prediction target (the number of hospital visits).
To estimate the variance in a patient's target value, we train multiple predictors over random data subsets, and infer prediction intervals for our algorithm's outputs~\cite{wager14}. The width of this interval is our estimate of the target's variance. Low variance means high prediction confidence.

We run a new \errorprofiling, with prediction
confidence as an explanatory attribute (\F\ref{fig:medical-bug-report-explained}).
Conditioning on prediction
confidence reduces the correlation in the full population.
For low confidence predictions, the correlation of error on age remains positive
and significant, but for high confidence users, the effect is
close to zero.
Similar results (omitted here) are obtained for users with an urgent-care
history.

\paragraph{Remediation.}
These results imply an immediate remediation strategy: when using this
algorithm to, say, tune insurance premiums, one should consider
the predictions' confidence.
For example, one might decide to automatically
tune premiums only for high-confidence
predictions.
This would result in about half of the users in our dataset
receiving customized premiums.
One could also develop a scheme that weighs any price
increase by prediction confidence.
\sysname can then be used to test either of
these approaches for disparate impact on the population.

\subsubsection{Movie Recommender}
\label{sec:movies}

This application highlights another interesting aspect about \sysname:
not all associations it discovers are true bugs that need remediation.
Some may simply be due to variations in \emph{utility} 
perceived by different users for some outcomes.

We train a movie recommender using the ALS
algorithm~\cite{matrixFactor} and MovieLens
data~\cite{movielens} (1M ratings provided by 6,040 users on a
total of 3,900 movies).
The ratings take values in $[1,5]$, and each user has rated at least
20 movies. 
The dataset includes user demographics (\eg age, gender)
and movie metadata (\eg release date, genre).
The system is trained to model the kinds of movies users generally
like, and it is configured to recommend $50$ new movies that a user has
not yet seen, but is likely to rate highly.

\paragraph{Detection.}
We test for disparities in the ``quality'' of movies
recommended to users. Let a movie's \emph{score}
be the average rating given to that movie by all users. We view a movie's score 
as an indicator of movie quality. We then use a \testing investigation to find associations
between the average score of a user's $50$ new recommended movies, and that
user's age or gender. \sysname's report (omitted for lack of space) shows
that men and older users receive recommendations for ``better'' movies 
than women and younger users respectively (correlation of $0.02$ for the gender disparity
and of $0.10$ for the age disparity).
This suggests that the recommendation system might provide higher utility to certain
user demographics, by recommending movies with higher scores.

\paragraph{Debugging.}
Although a low score indicates that a movie is generally disliked, it may be that
certain users actually enjoy movies regarded as ``bad'' by others. Indeed, certain movie genres (\eg horror movies or comedies) are consistently rated lower than others (\eg documentaries or dramas).
As we do not know the ratings that a user would give to the $50$ movies recommended by our system, 
we cannot directly assess the ``utility'' provided to each user.
However, we can test whether users that get recommendations for ``bad'' (low-scoring) movies are the ones that also watched ``bad'' movies in the past. In other words, we want to use \sysname's debugging capabilities to investigate
whether the discovered associations can be explained by some users' natural inclinations towards low-rated movies.

For each user, we compute the average score of all movies rated by that user. We then add a new attribute that splits users into two groups: those that watched movies with below (respectively above) average scores in the past. When added to \EA, i.e., treated as an \emph{explanatory attribute}, we find that a user's taste for low- or high-scoring movies accounts for the disparities found previously: After controlling for the scores of previously-rated movies, the difference in quality of recommended movies across gender is no longer significant.
Slight age-disparities remain, but the disparate effect is much lower than before (correlation of $0.04$).

\paragraph{Remediation.}
To conclude, we highlighted a situation in which apparent association bugs can be explained by simple and natural differences in user inclinations. We showed a successful use of \sysname's debugging features, in dismissing associations found by a previous \testing investigation. 
\begin{figure}[t]

\fbox{
\begin{minipage}{0.7\textwidth}
\ttfamily
\def\arraystretch{0.85}

\textbf{Report of associations of O=Labels on S\(\sb{i}\)=Race:}\\
\textbf{Global Population of size 1,324}\\[0.5em]
\textbf{* Labels associated with Race=Black:}\\[0.5em]
\begin{tabular}{@{}l c c c c@{}}
Label & Black & White & DIFF & p-value\\
 \midrule
\emph{Cart}    &      4\%   &  0\% &  [0.0137,0.0652] &  3.31e-05\\
\emph{Drum}     &     4\%   &  0\%  & [0.0095,0.0604]  & 3.83e-04\\
\emph{Helmet}   &     8\%   &  3\%  & [0.0096,0.0888] &  2.34e-03\\
\textbf{\emph{Cattle}}   &     \textbf{2\%}   &  \textbf{0\%}  & [0.0037,0.0432] &  4.73e-03
\end{tabular}
\vspace{1em}

\textbf{* Labels associated with Race=White:}\\[0.5em]
\begin{tabular}{@{}l c c c c@{}}
Label & Black & White & DIFF & p-value\\
 \midrule
\emph{Face Powder} &  \ 1\%  &  10\% & [-0.1339,-0.0525] & 5.60e-12\\
\emph{Maillot}     &  \ 4\%  &  15\% & [-0.1590,-0.0575] & 3.46e-10\\
\textbf{\emph{Person}}      & \textbf{96\%}  &  \textbf{99\%} & [-0.0563,-0.0042] & 6.06e-03\\
\emph{Lipstick}    &  \ 1\%  &   \ 4\% & [-0.0622,-0.0034] & 1.03e-02
\end{tabular}

\end{minipage}
}
\caption{ {\bf Racial Label Associations in the Image Tagger.}
Shows partial report of a \discovery(top\_k=35);
the four most strongly associated labels (for the binary difference metric DIFF) are shown for each race.
}
\label{fig:image-tagger-bug-report}
\end{figure}

\subsubsection{Image Tagger}
\label{sec:image-tagger}

We next show \sysname's \discovery capability from
the perspective of the developer of an image tagging system, who is willing to search for offensive labeling in racial groups.
To illustrate, we inspect the labels produced by
Caffe's~\cite{caffe} implementation of R-CNN~\cite{rcnn}, a ready-to-use image tagger, when applied to
photos of people from ImageNet~\cite{imagenet}.
The tagger was trained on images from ImageNet with $200$ tags, including images of people.
We tag $1{,}405$ images of black people and $1{,}243$
images of white people with $5$ labels each, and run a \discovery to find the 35 (top\_k) labels
most strongly associated with each race.
%\sysname automatically picks the regression metric to
%identify those labels.

\F\ref{fig:image-tagger-bug-report} shows \sysname's report.
It lists the labels most disparately applied to images of black people
(first table) and white people (second table); we show only 4 (of 35) labels
per race.
A developer could inspect all top\_k labels and
judge which ones deserve further scrutiny.
In \F\ref{fig:image-tagger-bug-report}, the `cattle' label might draw
attention due to its potentially negative connotation; upon
inspection, we find that none of the tagged images depict farm animals.
% The tagging of while people images with female clothing tags is also in
% most cases erroneous.
Moreover, black people receive the `person' tag less often, implying that the model is less accurate at detecting them.
Further work is needed to understand these errors.
While such analyses currently fall outside \sysname's scope, this example
shows that \sysname is effective at providing ``leads'' for investigations.
It will also help test the effectiveness of a remediation.

\begin{figure}[t]

\fbox{
\begin{minipage}{0.7\textwidth}
\ttfamily

\textbf{Report of associations of O=Income on S\(\sb{i}\)=Race}:\\
\textbf{Global Population of size 24,421}\\
p-value = 1.39e-53 ; NMI = [0.0063, 0.0139]\\[0.5em]
\begin{tabular}{@{}l | r | r | c | r | r@{}}
Income &   Asian &   Black & \dots &    White &      Total\\
\hline
<=50K &556(73\%)&\textbf{2061(88\%)}& &15647(75\%)&\textbf{18640 (76\%)}\\
 >50K &206(27\%)& 287(12\%)& & 5238(25\%)& 5781 (24\%)\\
\hline
Total &762 (3\%)&2348(10\%)& \dots &20885(86\%)&24421(100\%)
\end{tabular}
\vspace{	1em}

\textbf{1.\ Subpopulation of size 341}\\
\textbf{Context = Age <= 42, Hours <= 55, Job:\ Fed-gov}\\
p-value = 3.24e-03 ; NMI = [0.0085, 0.1310]\\[0.5em]
\begin{tabular}{@{}l | r | r | c | r | r@{}}
Income &   Asian &   Black & \dots &    White &      Total\\
\hline
<=50K &10(71\%)&\textbf{62(91\%)}&  &153(63\%)&\textbf{239 (70\%)}\\
>50K  & 4(29\%)& 6 (9\%)&  & 91(37\%)&102 (30\%)\\
\hline
Total &14 (4\%)&68(20\%)& \dots &244(72\%)&341(100\%)
\end{tabular}
\vspace{1em}

\textbf{2.\ Subpopulation of size 14,477}\\
\textbf{Context = Age <= 42, Hours <= 55}\\
p-value = 7.50e-31 ; NMI = [0.0070, 0.0187]\\[0.5em]
\begin{tabular}{@{}l | r | r | c | r | r@{}}
Income &   Asian &   Black & \dots &    White &      Total\\
\hline
<=50K &362(79\%)&\textbf{1408(93\%)}&  &10113(83\%)&\textbf{12157 (84\%)}\\
>50K  & 97(21\%)& 101 (7\%)&  & 2098(17\%)& 2320 (16\%)\\
\hline
Total &459 (3\%)&1509(10\%)& \dots &12211(84\%)&14477(100\%)
\end{tabular}
\end{minipage}
}

\caption{ {\bf Disparate Impact Reports on Race in the
Adult Income Dataset.} Shows the full population and two subpopulations with higher disparate effects.
}
\label{fig:adult-bug-report-race}
\end{figure}

\begin{figure}[t]
\fbox{
\begin{minipage}{0.7\textwidth}
\ttfamily

\parbox{\textwidth}{\textbf{Report of associations of O=Income on S\(\sb{i}\)=Gender}:}\\
\parbox{\textwidth}{\textbf{Global Population of size 24,421}}\\
\parbox{\textwidth}{p-value = 1.44e-178 ; NMI = [0.0381, 0.0540]}\\[0.5em]
\begin{tabular}{@{} l | r | r | r @{}}
Income &   Female &      Male &      Total\\
\hline
<=50K &\textbf{7218(89\%)}&\textbf{11422(70\%)}&18640 (76\%)\\
>50K  & 876(11\%)& 4905(30\%)& 5781 (24\%)\\
\hline
Total &8094(33\%)&16327(67\%)&24421(100\%)
\end{tabular}
\vspace{1em}

\textbf{1.\ Subpopulation of size 1,371}\\
\textbf{Context = 9 <= Education <= 11, Age >= 47}\\
p-value = 2.23e-35 ; NMI = [0.0529, 0.1442]\\[0.5em]
\begin{tabular}{@{} l | r | r | r @{}}
Income &   Female &      Male &      Total\\
\hline
<=50K &\textbf{423(88\%)}&\textbf{497(56\%)}& 920 (67\%) \\
>50K  & 57(12\%)&394(44\%)& 451 (33\%) \\
\hline
Total &480(35\%)&891(65\%)&1371(100\%)
\end{tabular}
\vspace{1em}

\textbf{2.\ Subpopulation of size 6,791}\\
\textbf{Context = Education >= 12}\\[-0.2em]
p-value = 3.71e-124 ; NMI = [0.0517, 0.0883]\\[0.5em]
\begin{tabular}{@{} l | r | r | r @{}}
Income &   Female &      Male &      Total\\
\hline
<=50K &\textbf{1594(76\%)}&\textbf{2156(46\%)}&3750 (55\%) \\
>50K  & 492(24\%)&2549(54\%)&3041 (45\%) \\
\hline
Total &2086(31\%)&4705(69\%)&6791(100\%)
\end{tabular}
\end{minipage}
}

\caption{ {\bf Disparate Impact Reports on Gender in the
Adult Income Dataset.} Shows the full population and two subpopulations with higher disparate effects.
}
\label{fig:adult-bug-report-gender}
\end{figure}

\subsubsection{Adult Income Census Dataset}
\label{sec:adult}

A second use case for \sysname is studying
discrimination in social datasets:
We use the Adult dataset~\cite{uci_repo}, which
reports census data and income (under or over \$50K) for 48,842
U.S. citizens.
Some disparate effects were found in prior algorithmic fairness works~\cite{zemel13,feldman14,hajian13,
kamishima12,zliobaite11,luong11}.
% Our goal here is to show how an analyst can use \sysname to
% further study this and other social datasets.

\F\ref{fig:adult-bug-report-race} and  \F\ref{fig:adult-bug-report-gender} show partial bug reports
for \testing for income biases on race (top) and gender (bottom).
We make three observations.
First, \sysname confirms previously known biases in the full dataset:
 $88\%$ of blacks have ${<}$\$50K-income
compared to $75\%$ of whites and $73\%$ of Asians.
Similarly, $89\%$ of women have low income compared to $70\%$ of men.

Second, \sysname reveals new insights into these biases.
For race (top), black people are strongly disfavored among people
younger than 42 working fewer than 55 hours a week
-- especially for federal government employees.
For gender (bottom), the groups where women are most disadvantaged are:
(1) older people with 9-11 years of education and (2)
(perhaps surprisingly) people with a higher education ($\geq$12 years of
education).
We are unaware of any prior works in the algorithmic fairness area that
have reported these particularly strong biases upon inspecting this dataset.

%Although this dataset has been studied profusely, we are unaware of
%prior reports of the strong race and gender %biases in these groups.
%We believe that the core ingredient missing from prior works is the efficient
%discovery of subpopulations of particularly high effect even if they are
%small in size.
%XXX  FLORIAN, is it true (have you seen any works that point these
%subpopulations out? Do they point out other subpopulations that we didn't
%find?) XXX
% [FT] None of the works I've seen that use this dataset actually give any epxlicit examples of discrimination contexts they find.

Third, as shown by the first context in \F\ref{fig:adult-bug-report-race}, \sysname
is capable of revealing even small contexts that show particularly strong
disparate effects.
%This leads to our third observation: as shown by the first context in
%\F\ref{fig:adult-bug-report}, \sysname is capable of revealing
%even small contexts that show
%particularly strong disparate effects.
% Since \sysname ensures simultaneous validity for all results in the
% report, the programmer can reliably compare these effect sizes, even
% for the small subpopulations.
%This capability has for instance attracted ACLU's attention, who wish to
%``zoom into'' their datasets to identify populations under particular
%distress.
% For developers, this feature might enable them to balance their economic
% interests with the desire not to discriminate, by exempting from data-driven
% decision-making those populations that appear most disadvantaged by it.
% (as Staples presumably did for Alaskans~\cite{staples}).

\section{Conclusion}
\label{sec:conclusion}

In an effort to rationalize and unify prior definitional foundations for algorithmic fairness, 
we have introduced {\em unwarranted associations}, a generic type of bug that characterizes disparate user treatment
in data-driven applications. To reason about these bugs, we have proposed the {\em UA framework and methodology} which uniquely combine multiple primitives and metrics with broad applicability, fine-grained exploration of biases in user subsets, incorporation of natural utility notions, and rigorous statistical assessments.

As an instantiation of our UA framework, we have designed and presented {\em \sysname}, 
a tool that helps responsible developers to thoroughly check data-driven
applications for unfair, discriminatory, or offensive user treatment.
Designed for ease-of-use, \sysname enables scalable,
statistically rigorous investigation of {\em unwarranted associations} between
application outcomes and protected user groups.
Our study of four applications shows the broad utility of \sysname's three
investigation types: \discovery of association bugs, \testing of suspected
bugs, and \errorprofiling.

We hope that our proposed tools will help developers answer recent calls from regulatory bodies, asking for data-driven algorithms to be tested and audited for fairness, in an effort to promote algorithmic accountability~\cite{algosAccountable}.   

\iffalse
We have presented \sysname, a tool to help privacy-conscious developers
 negotiate a world of increasingly complex application pipelines, machine learning algorithms, and data flows, with an ever-growing impact on users' lives. Designed for ease-of-use by developers, \sysname enables scalable, statistically rigorous investigation of unwarranted associations in data-driven applications. Our study of six real-world applications and datasets illustrates the broad utility of \sysname's three key investigation types: \discovery of association bugs, \testing of suspected bugs, and
\errorprofiling.
\fi

{
  \bibliographystyle{abbrv}
  \bibliography{paper}
}

\end{document}